\documentclass[12pt, draftclsnofoot, onecolumn,comsoc]{IEEEtran}

\usepackage{amsmath,graphicx}
\usepackage{amssymb}
\usepackage{subfigure}
\usepackage{graphicx}
\usepackage{graphics}
\usepackage{epstopdf}
\usepackage{amsmath}
\usepackage{multirow}
\usepackage{bm}
\usepackage{enumerate}
\usepackage{mathdots}
\usepackage{todonotes}
\usepackage{algorithmicx}
\usepackage{algpseudocode}
\usepackage{algorithm}
\usepackage{verbatim}
\usepackage{balance}
\usepackage{cite}
\usepackage{url}
\usepackage{booktabs}
\usepackage{tabularx}
\usepackage{acronym}

\newacro{ML}{Maximum Likelihood}
\newacro{LLF}{Log-Likelihood Function}
\newacro{SS-SW-OMP+Th}{Subcarrier Selection - Simultaneous Weighted - Orthogonal Matching Pursuit + Thresholding}
\newacro{SC-SS-SIMGW-IPM}{Spatially Consistent - Subcarrier Selection - Simultaneous Iterative Multi Gradient Weighted - Iterative Projection Maximization}
\newacro{SC-SS-SW-IPM}{Spatially Consistent - Subcarrier Selection - Simultaneous Weighted - Iterative Projection Maximization}
\newacro{AM}{Alternating Maximization}
\newacro{CSI}{Channel State Information}
\newacro{TTI}{Transmission Time Interval}
\newacro{AoA}{Angle-of-Arrival}
\newacro{AoD}{Angle-of-Departure}
\newacro{LSP}{Large-Scale Parameters}
\newacro{SSP}{Small-Scale Parameters}
\newacro{DS}{Delay Spread}
\newacro{AS}{Angular Spread}
\newacro{mmWave}{Millimeter Wave}
\newacro{LOS}{Line-of-Sight}
\newacro{NLOS}{Non Line-of-Sight}
\newacro{CRLB}{Cram\'{e}r-Rao Lower Bound}
\newacro{mmWave}{millimeter wave}
\newacro{MIMO}{Multiple-Input Multiple Output}
\newacro{NR}{New Radio}
\newacro{LoS}{Line-of-Sight}
\newacro{NLoS}{Non Line-of-Sight}
\newacro{CCDF}{Complementary Cumulative Distribution Function}
\newacro{V2X}{Vehicle-to-Everything}
\newacro{SVD}{Singular Value Decomposition}

\usepackage{color}
\definecolor{purple(x11)}{rgb}{0.63, 0.36, 0.94}
\definecolor{cadmiumgreen}{rgb}{0.0, 0.42, 0.24}

\newcommand{\diag}{\mathop{\mathrm{diag}}}
\newcommand{\trace}{\mathop{\mathrm{trace}}}

\usepackage{pifont}

\newtheorem{proof}{Proof}
\newcommand{\Nr}{N_{\mathrm{r}}}
\newcommand{\Nt}{N_{\mathrm{t}}}

\newcommand{\SNR}{\mathrm{SNR}}
\newcommand{\Qt}{Q_{\mathrm{t}}}
\newcommand{\Qr}{Q_{\mathrm{r}}}
\newcommand{\Ns}{N_{\mathrm{s}}}

\newcommand{\Ts}{T_{\mathrm{s}}}

\newcommand{\Gr}{G_{\mathrm{r}}}
\newcommand{\Gt}{G_{\mathrm{t}}}
\newcommand{\Lt}{L_{\mathrm{t}}}
\newcommand{\Lr}{L_{\mathrm{r}}}

\newcommand{\Nc}{N_{\mathrm{c}}}

\newcommand{\prc}{p_{\mathrm{rc}}}
\newcommand{\ar}{{\mathbf{a}}_{\mathrm{R}}}
\newcommand{\at}{{\mathbf{a}}_{\mathrm{T}}}

\newcommand{\AR}{{\mathbf{A}}_{\mathrm{R}}}
\newcommand{\AT}{{\mathbf{A}}_{\mathrm{T}}}

\newcommand{\jj}{{\mathrm{j}}}
\newcommand{\cH}{\mathbf{H}}

\newcommand{\bsfFrf}{{\bm{\mathsf{F}}_\text{RF}}}
\newcommand{\bsfFbb}{{\bm{\mathsf{F}}_\text{BB}}}

\newcommand{\bsfWrf}{{\bm{\mathsf{W}}_\text{RF}}}
\newcommand{\bsfWbb}{{\bm{\mathsf{W}}_\text{BB}}}

\newcommand{\bsfFrfher}{{\bm{\mathsf{F}}_\text{RF}^*}}
\newcommand{\bsfFbbher}{{\bm{\mathsf{F}}_\text{BB}^*}}

\newcommand{\bsfWrfher}{{\bm{\mathsf{W}}_\text{RF}^*}}
\newcommand{\bsfWbbher}{{\bm{\mathsf{W}}_\text{BB}^*}}

\newcommand{\be}{\begin{eqnarray}}
\newcommand{\ee}{\end{eqnarray}}

\def\ba{{\mathbf{a}}}

\def\bee{{\mathbf{e}}}

\def\bx{{\mathbf{x}}}

\def\b0{{\mathbf{0}}}

\def\bA{{\mathbf{A}}}

\def\bC{{\mathbf{C}}}

\def\bF{{\mathbf{F}}}
\def\bG{{\mathbf{G}}}
\def\bH{{\mathbf{H}}}
\def\bI{{\mathbf{I}}}

\def\bV{{\mathbf{V}}}

\def\bsfA{\bm{\mathsf{A}}}

\def\bsfC{\bm{\mathsf{C}}}

\def\bsfF{\bm{\mathsf{F}}}
\def\bsfG{\bm{\mathsf{G}}}
\def\bsfH{\bm{\mathsf{H}}}

\def\bsfP{\bm{\mathsf{P}}}
\def\bsfQ{\bm{\mathsf{Q}}}
\def\bsfR{\bm{\mathsf{R}}}
\def\bsfS{\bm{\mathsf{S}}}
\def\bsfT{\bm{\mathsf{T}}}
\def\bsfU{\bm{\mathsf{U}}}
\def\bsfV{\bm{\mathsf{V}}}
\def\bsfW{\bm{\mathsf{W}}}

\def\bsfY{\bm{\mathsf{Y}}}
\def\bsfZ{\bm{\mathsf{Z}}}

\def\sfj{{\mathsf{j}}}

\def\sfu{{\mathsf{u}}}

\def\sf0{{\mathsf{0}}}

\def\bsfa{{\bm{\mathsf{a}}}}

\def\bsfn{{\bm{\mathsf{n}}}}

\def\bsfs{{\bm{\mathsf{s}}}}

\def\bsfx{{\bm{\mathsf{x}}}}
\def\bsfy{{\bm{\mathsf{y}}}}

\def\bsf0{{\bm{\mathsf{0}}}}

\begin{document}
\title{Hybrid Precoding and Combining for Frequency-Selective mmWave MIMO Systems with Per-antenna Power Constraints}
\author{Javier Rodr\'{i}guez-Fern\'{a}ndez$^{\dag}$, Roberto L\'{o}pez-Valcarce$^{\ddag}$, and Nuria Gonz\'{a}lez-Prelcic$^{\dag}$
\\
$^\dag$ The University of Texas at Austin, Email: $\{$javi.rf,ngprelcic$\}$@utexas.edu \\
$^\ddag$ Universidade de Vigo, Email: valcarce@gts.uvigo.es}

\maketitle

\begin{abstract}
	Configuring hybrid precoders and combiners is a major challenge to deploy practical \ac{mmWave} communication systems. Prior work addresses the problem of designing hybrid precoders and combiner, yet focusing on finding solutions under a total transmit power constraint. The design of hybrid precoders and combiners in practical system, is constrained, however, by a per antenna transmit power, since each antenna element in the array is connected to a power amplifier (PA) that has to operate within its linear region. In this paper, we focus on the problem of hybrid precoding and combining with per-antenna power constraints, and under a frequency-selective bandlimited channel model. We first propose an all-digital solution to this problem, and develop a hybrid precoding and combining strategy that aims at matching this solution by minimizing the chordal distance between the all-digital precoders (combiners) and their hybrid approximations. Finally, since minimizing this metric does not guarantee that the final spectral efficiency will be maximized, we optimize the resulting spectral efficiency taking into account the per-antenna power constraints. Simulation results show the effectiveness of our all-digital and hybrid solutions, while emphasizing the differences with respect to the corresponding solution under a total power constraints. As shown in our numerical results, the proposed all-digital solution performs similarly to the case in which a total power constraint is considered. Further, our proposed hybrid solution is also shown to exhibit near-optimum performance, and the influence of different system parameters is also shown, thereby showing the suitability of our proposed framework to deploy practical \ac{mmWave} \ac{MIMO} systems.
 \end{abstract}


\section{Introduction}

Configuring antenna arrays is challenging in \ac{mmWave} frequency-selective scenarios. The main issues to be addressed are two-fold: i) obtaining high enough quality channel state information (CSI), which may be in terms of spatial covariance or channel estimates; and ii) jointly configuring hybrid architectures in both the analog and digital domains taking into account the hardware constraints \cite{Ahmed_MIMOprecoding_combining:CM2014}. Since part of the transmit and receive processing must be performed in the analog domain, accounting for hardware limitations, precoding and combining with hybrid architectures is significantly harder than its all-digital counterpart. Further, the analog precoder must accommodate different data streams for the different subcarriers using a small number of radio frequency (RF) chains. This results in a small number of degrees of freedom to design hybrid precoders and combiners, which makes it challenging to maximize the achievable spectral efficiency.

The idea of a hybrid analog-digital solution for the precoders and combiners under a total powe constraint was first proposed in \cite{Zhang_TSPHybrid}, and then developed in \cite{AyaRajAbuPiHea:Spatially-Sparse-Precoding:14} for sparse narrowband \ac{MIMO} channels at \ac{mmWave} frequencies. Perfect factorization of the precoders is obtained in \cite{Zhang_TSPHybrid}, but it requires a large number of RF chains, so that the solution is not feasible at \ac{mmWave} frequencies.  A common approach for the design of the \ac{mmWave} hybrid system is to factorize an approximation to the all-digital solution into the analog and the digital stages. Many methods have been proposed in the previous literature to solve this problem for the single-user and narrowband case \cite{GHP_SPAWC_2015}, \cite{RusRiaPreHea:Low-Complexity-Hybrid:15}, \cite{AAlkhateeb_Limited_Feedback}, yet the \ac{mmWave} channel is frequency-selective. 

Some recent papers consider the problem of hybrid precoding for wideband \ac{mmWave} systems, yet assuming a near-optimum sphere-decoding based combiner, e.g. \cite{Peng2017,Yu2016AlternatingMF,Dynamic_subarrays_Alkhateeb}. To the best of our knowledge, only \cite{Asilomar_2016_Jose},\cite{JSTSP_2018},\cite{Yu_2018}, \cite{SPAWC_2018} consider a more general approach for hybrid precoding and combining in \ac{mmWave} frequency-selective settings without this assumption about the combiner. In \cite{Asilomar_2016_Jose}, precoders and combiners are iteratively updated using several transmissions in the downlink (DL) and the uplink (UL) subject to a total power constraint for every subcarrier, thereby equally splitting the transmit power into each subcarrier. This results in optimizing the power allocation coefficients across the different data streams in an isolated manner for every subcarrier, so that the differences in \textit{effective} $\SNR$ for two data streams transmitted in two difference subcarriers are not exploited. Consequently, the final spectral efficiency is not fully optimized. Iteratively updating precoders and combiners dramatically increases the total training overhead, computational complexity, and may also lead to error propagation. The technique in \cite{JSTSP_2018}, also considering per-subcarrier total power constraints, has been shown to provide superior performance for \ac{mmWave} channels when the number of simulated multipath components is small and there is no clustering ($4$ multipath components for every user), yet at the expense of high computational complexity. In \cite{Yu_2018}, an Alternating-Minimization (AM)-based high-complexity iterative algorithm is proposed to find the hybrid precoders and combiners. The proposed \ac{AM} algorithm in \cite{Yu_2018} is numerically shown to converge to a stationary point that provides good performance, but it has not been proven to converge to the optimum solution when a realistic channel model is considered. Besides, the channel model in \cite{Yu_2018} is not bandlimited, i.e., it does not consider the practical effects of equivalent transmit-receive pulse-shaping plus analog filtering that may happen at the RF frontends, thereby increasing sparsity in the \ac{mmWave} \ac{MIMO} channel. Under this artifact, there is no power leakage in the extended virtual channel model representation \cite{RodGonVenHea:TWC_18}, which leads to the channels' singular basis being ''closer'' to each other in the Grassmanian manifold. Therefore, this makes it easier to design the hybrid precoders and combiners, since the analog precoder (combiner) can be designed to span a subspace that approximates a common singular subspace for the different subcarriers \cite{JSTSP_2018}. A common limitation of \cite{Asilomar_2016_Jose}, \cite{JSTSP_2018}, \cite{Yu_2018}, \cite{SPAWC_2018} is that they do not account for beam-squint effect, and \cite{Yu_2018} disregards the sampling rate of the \ac{mmWave} channel. In fact, by observing $(33)$ therein, one can realize that the \ac{mmWave} \ac{MIMO} channel is frequency-flat since its response is a delta function in the time domain. Further, \cite{Yu_2018} does not include any information as to the number of clusters, rays per cluster, angular spread, etc., used to generate the different channel realizations when simulating the proposed algorithm. This non-bandlimited model for the channel makes it easier to design hybrid precoders and combiners, and the proposed algorithm in \cite{Yu_2018} would likely exhibit significantly worse performance under practical channel models (i.e. as in 5G \ac{NR} channel model).

Finally, the technique in \cite{SPAWC_2018} is a greedy SVD-based low-complexity algorithm aiming at iteratively finding the RF and baseband precoders and combiners such that the Euclidean distances between the optimum unconstrained precoders and combiners and their hybrid counterparts are minimized. The common limitation of all these papers is that they deal with the problem of hybrid precoding and/or combining with a constraint on the total transmit power budget while equally splitting the power budget into every subcarrier. Considering per-antenna power constraints is, however, a more practical limitation to be considered, since the precoding/combing network includes a power amplifier per antenna, and each one of them has to operate in its linear range \cite{Globecom_16_per_antenna}. 

As to hybrid precoding with per-antenna power constraints, to the best of our knowledge, only \cite{Precoding_Massive_MIMO_subarrays}, and \cite{Globecom_16_per_antenna} deal with this problem for the narrowband scenario. A downlink massive MISO setting is considered in \cite{Precoding_Massive_MIMO_subarrays}, and the problem of maximizing the mutual information is formulated. The analog precoder is iteratively designed based on a per-antenna power allocation solution only, without considering the power additivity coming from having several RF chains. The baseband precoder, conversely, is designed subject to the RF chain power constraints only, whereby an iterative power allocation algorithm for the digital precoder is proposed. In our previous work presented in \cite{Globecom_16_per_antenna}, the problem of maximizing the spectral efficiency with per antenna power constraints in a \ac{mmWave} \ac{MIMO} system is considered. A convex relaxation of the original problem is performed, such that closed-form suboptimal solutions for both the all-digital and hybrid precoders can be derived. Furthermore, we propose another suboptimal solution for the original problem, which is in fact an upper bound to the original problem that improves performance when compared to the relaxed solution.

In this paper, we develop the first solution for hybrid precoders and combiners with per-antenna power constraints for frequency-selective \ac{mmWave} \ac{MIMO} systems. The RF analog processing is implemented using a fully-connected phase-shifters-based architecture, where each signal in the digital domain is fed into every RF chain. Whereas most of previous hybrid transceiver designs consider the maximization of the mutual information, thus decoupling the designs of the precoder and combiner, and with a total power constraint, we address the problem of maximizing the spectral efficiency by jointly designing the precoder and combiner under per-antenna constraints. To perform this task, we first consider an all-digital design without any hardware-related constraints, and we optimize the spectral efficiency under the per-antenna power constraints. Thereby, some insight into this problem can be obtained and a strict upper bound for the maximum achievable performance can be also derived. Since the resulting problem is difficult, we introduce a convex relaxation leading to a suboptimal solution that can be obtained in closed form. 

The contributions of this paper are summarized hereafter:
\begin{itemize}
	\item We formulate the problem of finding all-digital precoders and combiners maximizing the spectral efficiency of a \ac{MIMO}-OFDM/SC-FDE system with per-antenna power constraints. Owing to the non-convex nature of the problem, we propose a convex relaxation of the original problem and find its optimal solution. Further, we also propose another solution and prove that it is in fact an upper bound for the original problem.
	\item Inspired by the all-digital solution and the discussion in \cite{AyaRajAbuPiHea:Spatially-Sparse-Precoding:14}, we propose a design criterion for hybrid precoders and combiners based on minimizing the chordal distance between the all-digital solutions and their hybrid counterparts. 
	\item We propose a design method for the analog precoder subject to the hardware constraints of the phase shifting network. This method is focused on finding a frequency-flat unconstrained precoder that, on average, best matches the singular subspaces of the different subchannels, thereby minimizing the subspace distance between the analog precoder (combiner) and the all-digital spatial filters. Further, we prove that optimizing this metric leads to optimizing the lower bound of the subspace distance between the all-digital precoders (combiners) and their hybrid counterparts. Finally, we design the baseband precoders and combiners subject to the per-antenna power constraints. A major advantage of our proposed hybrid precoding and combining strategy is that we find closed-form solutions for both the RF precoder (combiner) and the singular subspaces of the baseband precoders (combiners). The singular values of the baseband precoders are obtained by optimizing the resulting spectral efficiency, which is a convex problem that can be efficiently solved.
	\item We evaluate and compare the performance of the all-digital and hybrid solutions under both perfect CSI and channel estimates using realistic small-scale parameters directly taken from the 5G \ac{NR} channel model defined in TR 38.901 \cite{5G_channel_model}. For the channel estimates, we consider the Subcarrier-Selection Simultaneous-Weighted Orthogonal Matching Pursuit + Thresholding (SS-SW-OMP+Th) in \cite{RodGonVenHea:TWC_18} for \ac{mmWave} channels in the absence of beam-squint, which, under certain assumptions on the \ac{mmWave} \ac{MIMO} channel, has been shown to be asymptotically efficient. Further, we also consider its generalization to deal with frequency-selective antenna arrays in \cite{Globecom_18_beam_squint}, and show the dependence of the spectral efficiency on the system's bandwidth, which directly relates to the dependence between the channels' singular subspaces for the different subcarriers. Our proposed all-digital solution with per-antenna constraints is shown to perform similarly to the design considering a total power constraint. Finally, we also provide some insight into why the performance gap is smal with respect to the design with a total power constraint.
\end{itemize}

\textbf{Notation}: We use the following notation throughout this paper: bold uppercase $\bA$ is used to denote matrices, bold lowercase $\ba$ denotes a column vector and non-bold lowercase $a$ denotes a scalar value. We use ${\cal A}$ to denote a set. Further, $\|\bA\|_F$ is the Frobenius norm of a matrix $\bA$. The symbols $\bA^*$, $\bA^\text{C}$, $\bA^T$ and $\bA^\dag$ denote the conjugate transpose, conjugate, transpose and Moore-Penrose pseudoinverse of a matrix $\bA$, respectively. The $(i,j)$-th entry of a matrix $\bA$ is denoted using $\{\bA\}_{i,j}$. Similarly, the $i$-th entry of a column vector $\ba$ is denoted as $\{\ba\}_i$. The identity matrix of order $N$ is denoted as $\bI_N$. We use ${\cal CN}(\mathbf{\mu},\bC)$ to denote a circularly symmetric complex Gaussian random vector with mean $\mathbf{\mu}$ and covariance matrix $\bC$. We use $\mathbb{E}$ to denote expectation. Discrete-time signals are represented as $\bx[n]$, while frequency-domain signals are denoted using $\bsfx[k]$.

\section{System model}
\label{section:systemmodel}
\begin{figure*}[t!]
\centering
\includegraphics[width=\textwidth]{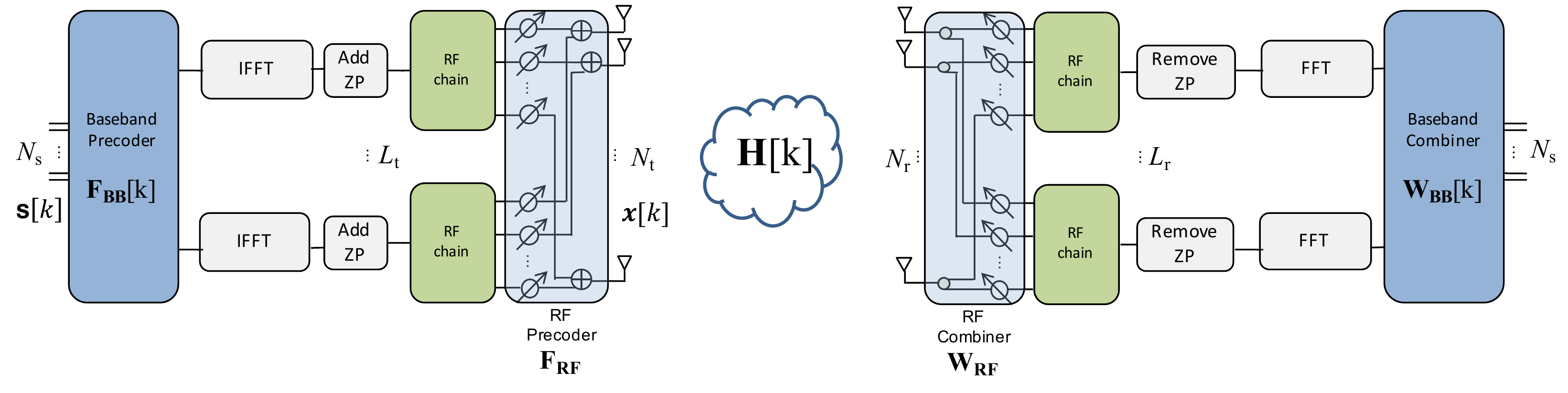} 
\caption{Illustration of the structure of a hybrid MIMO architecture, which include analog and digital precoders and combiners.}     
\label{fig:hybrid_architecture}        
\end{figure*}

We consider a single-user OFDM based hybrid \ac{mmWave} \ac{MIMO} link employing $K$ subcarriers to send a block of $N_\text{OFDM}$ data symbol streams $\bsfs_t[k] \in \mathbb{C}^{\Ns \times 1}$, $t = 0,\ldots,N_\text{OFDM}$, with $\mathbb{E}\{\bsfs_t[k] \bsfs_t^*[k]\} = \frac{1}{\Ns} \bI_{\Ns}$, using a transmitter with $\Nt$ antennas and a receiver with $\Nr$ antennas. The system is based on a hybrid \ac{MIMO} architecture as shown in Fig. \ref{fig:hybrid_architecture}, with $L_\text{t}$ and $L_\text{r}$ RF chains at the transmitter and receiver sides. At the transmitter side, a frequency-selective hybrid precoder $\bsfF[k] \in \mathbb{C}^{\Nt \times \Ns}$ is used, with $\bsfF[k] = \bsfFrf \bsfFbb[k]$, $k=0,\ldots,K-1$, where $\bsfFrf$ is the analog precoder and $\bsfFbb[k]$ its digital counterpart. The RF precoder and combiner are implemented using a fully-connected network of phase shifters, as described in \cite{RiaRusPreAlkHea:Hybrid-MIMO-Architectures:16}. Notice that the analog precoder is frequency-flat, while the digital precoder may be different for every subcarrier. The symbol blocks are transformed into the time domain using $L_\text{t}$ parallel $K$-point IFFTs. 
As in \cite{LarThoCud:Air-interface-design-and-ray-tracing:13}, we consider $Z_\text{CP}$-length Zero-Padding (ZP) to both suppress Inter Block Interference (IBI) and account for the RF circuitry reconfiguration time. The time domain transmitted signal can thereby be expressed as
\begin{equation}
	\bx[n] = \bsfF_\text{RF} \frac{1}{K}\sum_{t=0}^{N_\text{OFDM}-1}\sum_{k=0}^{K-1} \bsfF_\text{BB}[k] \bsfs_t[k] e^{\sfj \frac{2\pi k (n - Z_\text{CP} - t(K + Z_\text{CP}))}{K}} w_K[n - t(K + Z_\text{CP})],
	\label{equation:tx_signal}
\end{equation}
for $n = tZ_\text{CP},\ldots,tZ_\text{CP}+K-1$, and $\bx[n] = \bm 0$ for $n = t(Z_\text{CP}-1),\ldots,t(Z_\text{CP}-1) + Z_\text{CP}-1$, and $w_N[n]$ is the $N$-length rectangular pulse $w_N[n] = 1$ for $n = 0,\ldots,N-1$, and $w_N[n] = 0$ otherwise.

The \ac{MIMO} channel between the transmitter and the receiver is assumed to be frequency-selective, having a delay tap length $\Nc$ in the time domain.
Let $\rho_\text{L}$ be the pathloss between transmitter and receiver, $C$, $R_c$ be the number of clusters and rays within $c$-th cluster, $\Ts$ be the sampling period, $\prc(\tau)$ be a filter including the effects of pulse-shaping and other analog filtering evaluated at $\tau$, $\alpha_{c,r} \in {\mathbb{C}}$ be the complex gain of the $(c,r)$-th path, $\tau_{c,r} \in {\mathbb{R}}$ be the delay of the $(c,r)$-th path, $\phi_{c,r} \in [0, 2\pi)$ and $\theta_{c,r} \in [0, 2\pi)$ be the angles-of-arrival and departure (AoA/AoD) of the $(c,r)$-th path, and $\ar(\phi_{c,r}) \in {\mathbb{C}}^{\Nr\times1}$, $\at(\theta_{c,r}) \in {\mathbb{C}}^{\Nt\times1}$ be the array steering vectors for the receive and transmit antennas. Then, the $d$-th delay tap of the channel is represented
by a $\Nr \times \Nt$ matrix denoted as $\cH_d$, $d=0,\ldots,\Nc-1$, which, assuming a geometric channel model \cite{schniter_sparseway:2014}, can be written as
\be
\hspace*{-3.5mm}\cH_d &=\hspace*{-1mm}& \sqrt{\frac{N_\text{t} N_\text{r}}{\rho_\text{L}\sum_{c=1}^{C}R_c}}\sum_{c = 1}^{C}\sum_{r=1}^{R_c}\alpha_{c,r}\prc(dT_\text{s}-\tau_{c,r})\ar(\phi_{c,r})\at^*(\theta_{c,r}), \label{eqn:channel_model}
\ee
Every matrix in \eqref{eqn:channel_model} can be written in a more compact way as
\be
\cH_d &=& \AR \bG_d\AT^*, \label{eqn:channel_compact}
\ee 
where $\bG_d \in {\mathbb{C}}^{\sum_{c=1}^{C}R_c\times \sum_{c=1}^{C}R_c}$ is diagonal with non-zero complex entries, and $\AR \in {\mathbb{C}}^{\Nr\times \sum_{c=1}^{C}R_c}$ and $\AT \in {\mathbb{C}}^{\Nt\times \sum_{c=1}^{C}R_c}$ contain the receive and transmit array steering vectors $\ar(\phi_{c,r})$ and $\at(\theta_{c,r})$, respectively. 

Finally, the frequency-domain \ac{MIMO} channel matrix at subcarrier $k$ can be written in terms of the different delay taps as
\be
\bsfH[k]=\sum_{d=0}^{\Nc-1} \cH_d e^{-\jj\frac{2\pi k}{K}d} = \bsfA_\text{R} \bsfG[k] \bsfA_\text{T}^*.
\label{equation_channel_k_perf}
\ee
In the presence of beam-squint, the array response vectors are frequency-dependent, and thus the channel matrix in \eqref{equation_channel_k_perf} can be accordingly modified as \cite{Globecom_18_beam_squint}, \cite{Akbar:BeamSquint}
\be
\bsfH[k] = \bsfA_\text{R}[k] \bsfG[k] \bsfA_\text{T}^*[k].
\label{equation_channel_k_perf_beam_squint}
\ee
Assuming that the receiver applies a hybrid combiner $\bsfW[k] = \bsfWrf \bsfWbb[k] \in {\mathbb{C}}^{\Nr\times N_\text{s}}$, the received signal at subcarrier $k$ can be written as
\be
\begin{split}
\bsfy[k]  =  \bsfWbbher[k] \bsfWrfher \bsfH[k] \bsfFrf \bsfFbb[k] \bsfs[k]\\
 + \bsfWbbher[k] \bsfWrfher \bsfn[k],\\
\end{split}
\label{equation:signal_model}
\ee
where $\bsfn[k] \sim \mathcal{CN}\left(0,\sigma^2 \mathbf{I}\right)$  is the circularly symmetric complex Gaussian distributed additive noise vector.

\section{Problem formulation}

We assume perfect channel state information (CSI) at the transmitter and receiver, and focus on the problem of designing hybrid precoders and combiners maximizing the spectral efficiency (or, equivalently, the achievable rate), subject to per-antenna power constraints, as in \cite{Globecom_16_per_antenna}. Let us define the receive $\SNR$ at the receive antenna level as $\SNR \triangleq P_\text{tx}/\sigma^2$, with $P_\text{tx}$ being the transmitted power, and $\sigma^2$ the receive noise variance. Under transmitted Gaussian signaling, the spectral efficiency achieved when transmitter and receiver use a precoder $\bsfF[k]$ and a combiner $\bsfW[k]$ can be expressed as \cite{MIMO_capacity}
\begin{equation}
	\underset{k = 0,\ldots,K-1}{{\cal R}(\bsfF[k], \bsfW[k])} = \frac{1}{K}\sum_{k=0}^{K-1}{\log_2\left|\bI_{\Ns} + \frac{\SNR}{\Ns} (\bsfW^*[k]\bsfW[k])^{-1}\bsfW^*[k]\bsfH[k]\bsfF[k]\bsfF^*[k]\bsfH^*[k]\bsfW[k]\right|}.
	\label{equation:spectral_efficiency}
\end{equation}
By taking the $K$-point DFT of $\bx[n]$ in \eqref{equation:tx_signal} after discarding the Zero-Prefix, and using Parseval's theorem, the power constraint for the $t$-th OFDM symbol forwarded through the $j$-th transmit antenna, $\bsfx_{j,t}[k] = [\bsfF[k]]_{j,:} \bsfs_t[k]$, is given by
\begin{equation}
\begin{split}
	\sum_{k=0}^{K-1}\mathbb{E}\{|\bsfx_j[k]|^2\} \leq p_j, \qquad & j = 1,\ldots,\Nt, \\
\end{split}
\end{equation}
so that developing the left-handed term for the $t$-th transmitted OFDM symbol allows us to write
\begin{equation}
\begin{split}
	\sum_{k=0}^{K-1}\mathbb{E}\{|\bsfx_{j,t}[k]|^2\} &= \sum_{k=0}^{K-1}\trace\{\left[\bsfF[k]\right]_{j,:}\underbrace{\mathbb{E}\{\bsfs_t[k]\bsfs_t^*[k]\}}_{\frac{1}{\Ns}\bI_{\Ns}}\left[\bsfF[k]\right]_{j,:}^*\} \\
	&= \frac{1}{\Ns}\sum_{k=0}^{K-1}\|\left[\bsfF[k]\right]_{:,j}\|_2^2,
\end{split}
\end{equation}
which is essentially a constraint over the $j$-th row of the frequency-selective precoder. Therefore, the power constraint for the $j$-th transmit antenna can be expressed in matrix form as
\begin{equation}
\begin{split}
	\frac{1}{\Ns}\bee_j^*\left(\sum_{k=0}^{K-1}\bsfF[k] \bsfF^*[k]\right) \bee_j \leq p_j, \qquad & j = 1,\ldots,\Nt. \\
\end{split}
	\label{equation:power_constraints}
\end{equation}
In \eqref{equation:power_constraints}, $\bee_j \in \mathbb{C}^{\Nt \times 1}$ is the $j$-th element of the canonical vector basis, and $p_j > 0$ is the maximum allowable average transmit power at the $j$-th transmit antenna. Besides the constraints in \eqref{equation:power_constraints}, additional hardware constraints imposed by the analog precoding stage need also be taken into account. We consider that transmitter and receiver employ fully-connected hybrid architectures, as shown in Fig. \ref{fig:hybrid_architecture}. Let ${\cal M}^{\Nt \times \Lt}(\Qt)$ and ${\cal M}^{\Nr \times \Lr}(\Qr)$ denote the sets of matrices with entries having unit magnitude and phase taken from a discrete set of values, corresponding to $\Qt$ and $\Qr$ quantization bits, respectively. Then, the analog precoder and combiner must also satisfy $\bsfFrf \in {\cal M}^{\Nt \times \Lt}(\Qt)$ and $\bsfWrf \in {\cal M}^{\Nr \times \Lr}(\Qr)$.
Since incorporating the hardware-specific constraints into an optimization problem results in a non-convex problem, we will first remove them to gain some insight into the design of hybrid precoders and combiners with per-antenna power constraints. Thereafter, we will focus on finding the frequecy-selective hybrid approximations that are best matched to the all-digital solution in the sense of minimizing their respective chordal distances.

\section{All-digital design}
\label{section:all-digital}
In this section, we assume an all-digital implementation of the precoder and combiner, and consider the problem of jointly maximizing the spectral efficiency ${\cal R}(\bsfF[k], \bsfW[k])$ subject to the per-antenna power constraints. This problem can be stated as
\begin{equation}
	\underset{\underset{k = 0,\ldots,K-1}{\bsfF[k],\bsfW[k]}}{\max}\,{\cal R}(\bsfF[k],\bsfW[k]) \qquad \text{subject to} \qquad \frac{1}{\Ns}\bee_j^*\left(\sum_{k=0}^{K-1}\bsfF[k] \bsfF^*[k]\right) \bee_j \leq p_j, \quad j = 1,\ldots,\Nt.
	\label{equation:opt_se_original}
\end{equation}
The closed-form solution to the problem in \eqref{equation:opt_se_original} is unknown even for the narrowband scenario \cite{Globecom_16_per_antenna}. In \cite{PiOptimal}, however, the problem of maximizing the mutual information was solved for the narrowband scenario using an iterative approach based on Newton's method. In this paper, we follow the philosophy in \cite{Globecom_16_per_antenna} and extend its formulation to the frequency-selective scenario. 

In \cite{Globecom_16_per_antenna}, we proved that, in the narrowband scenario, the optimal precoder must satisfy all per-antenna power constraints with equality. For the frequency-selective scenario, this result can be extended as:

\textbf{Lemma 1.} \textit{Consider the problem}
\begin{equation}
\begin{split}
	\left\{\{\bsfF^\star[k]\}_{k=0}^{K-1},\{\bsfW^\star[k]\}_{k=0}^{K-1}\right\} = &\underset{\{\bsfF[k]\}_{k=0}^{K-1},\{\bsfW[k]\}_{k=0}^{K-1}}{\arg\,\max}\,{\cal R}(\bsfF[k],\bsfW[k]) \\
	& \text{subject to} \qquad \frac{1}{\Ns}\bee_j^*\left(\sum_{k=0}^{K-1}\bsfF[k]\bsfF^*[k]\right)\bee_j \leq p_j, \quad j = 1,\ldots,\Nt.
\end{split}
\label{equation:opt_lemma}
\end{equation}
\textit{Then, the optimal precoder for \eqref{equation:opt_lemma} satisfies all the per-antenna power constraints with equality.}

\begin{proof} Consider the maximization of the spectral efficiency in \eqref{equation:spectral_efficiency} and let $\{\bsfW^\star[k]\}_{k=0}^{K-1}$ be the hybrid combiner maximizing \eqref{equation:opt_lemma}. Then, consider the SVD $\bsfW^\star[k] = \bsfU_W^\star[k] \bm \Sigma_W^\star[k] (\bsfV_W^{\star})^*[k]$. If we consider the optimization of the spectral efficiency as a function of $\bsfF[k]$, $k = 0,\ldots,K-1$, the received noise covariance is given by $\bsfV_W^\star[k]\bm \Sigma_W^2[k] (\bsfV_W^\star[k])^*$. Therefore, the spectral efficiency is only a function of $\bsfU_W^\star[k]$, $\bsfF[k]$:
\begin{equation}
	{\cal R}(\bsfF[k], \bsfW^\star[k]) = \frac{1}{K}\sum_{k=0}^{K-1}\log_2\left|\bI_{\Ns} + \frac{\SNR}{\Ns} \bsfF^*[k] \bsfH^*[k] \bsfU_W^\star[k] (\bsfU_W^\star[k])^* \bsfH[k] \bsfF[k]\right|.
	\label{equation:mutual_info_lemma1}
\end{equation}
The function in \eqref{equation:mutual_info_lemma1} is the mutual information achieved by the precoder $\bsfF[k]$ over a frequency-selective channel $\tilde{\bsfH}[k] = \left(\bsfU_W^\star[k]\right)^*\bsfH[k]$. Since the hybrid precoders and combiners can be independently designed for each subcarrier, we may follow the same reasoning as in \cite{Globecom_16_per_antenna}. Thus, using Lemma 1 in \cite{Globecom_16_per_antenna}, the optimum solution for a given subcarrier satisfies the per-antenna power constraints with equality, and this holds for every subcarrier. Thereby, the power constraints are met with equality for the optimum solution. This completes the proof.
\end{proof}

Beyond Lemma 1, characterizing the solution to problem \eqref{equation:opt_se_original} is difficult. Due to this, we resort to a suboptimal yet tractable approximation to solve this optimization problem. To do this, let us consider the left-handed term of the $j$-th constraint in \eqref{equation:opt_se_original}. This term can be rewritten as
\begin{equation}
	\frac{1}{\Ns} \bee_j^*\left(\sum_{k=0}^{K-1}{\bsfF[k]\bsfF^*[k]}\right)\bee_j \leq \frac{1}{\Ns}\sum_{k=0}^{K-1}\sigma_\text{max}^2\left(\bsfF[k]\right),
\end{equation}
where $\sigma_\text{max}(\bA)$ denotes the largest singular value of the matrix $\bA$. Now, let us define $p_0 \triangleq \underset{j}{\min}\,p_j$. Therefore, the set of precoders satisfying $\frac{1}{\Ns}\sum_{k=0}^{K-1}{\sigma_\text{max}^2\left(\bsfF[k]\right)} \leq p_0$ is feasible for problem \eqref{equation:opt_se_original}, although this choice will be suboptimal, in general. In view of this, the original problem can be relaxed as follows:
\begin{equation}
\begin{split}
	\left\{\{\bsfF^\star[k]\}_{k=0}^{K-1},\{\bsfW^\star[k]\}_{k=0}^{K-1}\right\} = &\underset{\{\bsfF[k]\}_{k=0}^{K-1},\{\bsfW[k]\}_{k=0}^{K-1}}{\arg\,\max}\,{\cal R}(\bsfF[k],\bsfW[k]) \\
	& \text{subject to} \qquad \frac{1}{\Ns}\sum_{k=0}^{K-1}\sigma_\text{max}^2\left(\bsfF[k]\right) \leq p_0,
\end{split}
	\label{equation:opt_se_relaxed}
\end{equation}
which has an optimal solution that will be found in the next paragraphs.

The solution to the problem \eqref{equation:opt_se_relaxed} can be characterized by the following result:

\textbf{Theorem 1.} \textit{Let us consider the \ac{SVD} of the rank-$\Ns$ approximation of the \ac{MIMO} channel $\bsfH[k]$ as $\bsfH[k] \approx \tilde{\bsfU}_H[k] \tilde{\bm \Sigma}_H[k] \tilde{\bsfV}_H^*[k]$, with $\tilde{\bsfU}_H[k] \in \mathbb{C}^{\Nt \times \Ns}$ and $\tilde{\bsfV}_H[k] \in \mathbb{C}^{\Nr \times \Ns}$ respectively comprise the $\Ns$ left and right singular vectors of the channel matrix $\bsfH[k]$ corresponding to the $\Ns$ largest singular values, which are contained in the diagonal matrix $\tilde{\bm \Sigma}_H[k] \in \mathbb{C}^{\Ns \times \Ns}$. Then, the solution to problem \eqref{equation:opt_se_relaxed} is given by
\begin{equation}
	\bsfF[k] = \tilde{\bsfV}_H[k] \bsfQ_F[k] \bm \Sigma_F^2[k], \qquad \bsfW[k] = \tilde{\bsfU}_H[k] \bsfR_W[k], \qquad \tilde{\bm \Sigma}_F[k] = \rho[k]\bI_{\Ns}
\end{equation}
with $\bsfQ_F[k] \in \mathbb{C}^{\Ns \times \Ns}$ any unitary matrix, and $\bsfR_W[k] \in \mathbb{C}^{\Ns \times \Ns}$ any invertible matrix. The coefficients $\rho[k]$ are the solution to the following optimization problem
\begin{equation}
\begin{split}
	{\cal R}(\bsfF[k],\bsfW[k]) &= \frac{1}{K}\sum_{k=0}^{K-1}\log_2\left|\bI_{\Ns} + \frac{\rho^2[k] \SNR}{\Ns} \tilde{\bm \Sigma}_H^2[k] \right|, \\
	&\text{subject to} \qquad \frac{1}{\Ns}\sum_{k=0}^{K-1}\rho^2[k] \leq p_0.
\end{split}
\label{equation:opt_se_relaxed_Th}
\end{equation}
The optimization problem in \eqref{equation:opt_se_relaxed_Th} comprises a linear (convex) combination of log-concave functions with a linear constraint on the power allocation scalars $\rho^2[k]$. Since the constraint in \eqref{equation:opt_se_relaxed_Th} is convex, and the objective function is also convex (concave), the problem in \eqref{equation:opt_se_relaxed_Th} is convex. Thus, it can be solved using any convex optimization tool.}

\begin{proof} Let us consider the SVDs of the frequency-selective precoders and combiners,
\begin{equation}
	\bsfF[k] = \bsfU_F[k] \bm \Sigma_F[k] \bsfV_F^*[k], \qquad \bsfW[k] = \bsfU_W[k] \bm \Sigma_W[k] \bsfV_W^*[k],
\end{equation}
with $\bsfU_F[k] \in \mathbb{C}^{\Nt \times \Ns}$, $\bm \Sigma_F[k] \in \mathbb{C}^{\Ns \times \Ns}$, $\bsfV_F[k] \in \mathbb{C}^{\Ns \times \Ns}$, $\bsfU_W[k] \in \mathbb{C}^{\Nr \times \Ns}$, $\bm \Sigma_W[k] \in \mathbb{C}^{\Ns \times \Ns}$, $\bsfV_W[k] \in \mathbb{C}^{\Ns \times \Ns}$. The term $\left(\bsfW^*[k] \bsfW[k]\right)^{-1}$ in \eqref{equation:spectral_efficiency} is given by $\left(\bsfW^*[k] \bsfW[k]\right) = \bsfV_W[k] \bm \Sigma_W^{-2}[k] \bsfV_W^*[k]$ and the spectral efficiency is found to only be a function of $\bsfU_F[k] \bm \Sigma_F[k] \bsfU_W[k]$:
\begin{equation}
	{\cal R}(\bsfF[k],\bsfW[k]) = \frac{1}{K}\sum_{k=0}^{K-1}\log_2\left|\bI_{\Ns} + \frac{\SNR}{\Ns} \bsfU_W^*[k] \bsfH[k] \bsfU_F[k] \bm \Sigma_F^2[k] \bsfU_F^*[k] \bsfH[k] \bsfU_W[k] \right|.
	\label{equation:opt_se_Th1_SVD}
\end{equation}
Thus, since $\bsfU_F[k]$, $\bsfU_W[k]$ are unitary and $\bm \Sigma_F[k]$ is diagonal, we have to maximize \eqref{equation:opt_se_Th1_SVD} subject to
\begin{equation}
	\bsfU_F^*[k]\bsfU_F[k] = \bI_{\Ns}, \qquad \bsfU_W^*[k]\bsfU_W[k] = \bI_{\Ns}, \qquad \frac{1}{\Ns}\sum_{k=0}^{K-1} \sigma_\text{max}^2\left(\bsfF[k]\right) \leq p_0.
	\label{equation:constraints_Th1}
\end{equation}
The main source of difficulty in solving \eqref{equation:opt_se_Th1_SVD} comes from the third constraint in \eqref{equation:constraints_Th1}. Even with the knowledge of the channel matrix $\bsfH[k]$, splitting the power budget $p_0$ into the different $\bm \Sigma_F[k]$ is difficult. Due to this, we will focus on finding the optimum semi-unitary matrices $\bsfU_F[k]$, $\bsfU_W[k]$ that maximize \eqref{equation:opt_se_Th1_SVD}. If we use Sylvester's theorem, we may express \eqref{equation:opt_se_Th1_SVD} as
\begin{equation}
	{\cal R}(\bsfF[k],\bsfW[k]) = \frac{1}{K}\sum_{k=0}^{K-1}\log_2\left|\bI_{\Ns}+ \frac{\SNR}{\Ns} \bsfU_W[k] \bsfU_W^*[k] \bsfH[k] \bsfU_F[K] \bm \Sigma_F^2[k] \bsfU_F^*[k] \bsfH^*[k] \right|.
	\label{equation:opt_se_Th1_SVD_2}
\end{equation}
Let us define a positive semidefinite Hermitian matrix $\bsfC[k] \triangleq \frac{\SNR}{\Ns} \bsfH[k] \bsfU_F[k] \bm \Sigma_F^2[k] \bsfU_F^*[k] \bsfH[k]$. Then, using Lemma 3 from \cite{Witsenhausen}, and the fact that the $\log(\cdot)$ function is log-concave,
\begin{equation}
\begin{split}
	{\cal R}(\bsfF[k],\bsfW[k]) &= \frac{1}{K} \sum_{k=0}^{K-1}\log_2\left|\bI_{\Ns} + \bsfU_W[k] \bsfU_W^*[k] \bsfC[k] \right| \\ 
	& \leq \frac{1}{K} \sum_{k=0}^{K-1}\log_2 \left(\prod_{i=1}^{\Nr} \left(1 + \lambda_i(\bsfU_W[k] \bsfU_W^*[k])\lambda_i(\bsfC[k])\right)\right) \\
	&= \frac{1}{K} \sum_{k=0}^{K-1}\log_2 \left(\prod_{i=1}^{\Ns} (1 + \lambda_i(\bsfC[k]))\right),
\end{split}
\label{equation:Witsenhausen_Uw}
\end{equation}
where the second step follows from the fact that $\bsfU_W[k] \bsfU_W^*[k]$ has eigenvalues $\lambda_i(\bsfU_W[k] \bsfU_W^*[k]) = 1$ for $1 \leq i \leq \Ns$ and zero otherwise. The upper bound in \eqref{equation:Witsenhausen_Uw} is achieved if $\bsfU_W[k]\bsfU_W^*[k] \bsfC[k]$ commute, and since there is no constraint on the relation between the different $\bsfU_W[k]$ beyond being semi-unitary, this applies to every $0 \leq k \leq K-1$. Let us consider an eigendecomposition of $\bsfC[k]$ given by $\bsfC[k] = \bsfU_C[k] \bm \Lambda[k] \bsfU_C^*[k]$, with $\bsfU_C[k] \in \mathbb{C}^{\Nt \times \Ns}$, and $\bm \Lambda_C[k] \in \mathbb{C}^{\Ns \times \Ns}$. Then, the upper bound in \eqref{equation:Witsenhausen_Uw} requires that
\begin{equation}
	\bsfU_W[k] \bsfU_W^*[k] \bsfU_C[k] \bm \Lambda_C[k] \bsfU_C^*[k] = \bsfU_C[k] \bm \Lambda_C[k] \bsfU_C^*[k] \bsfU_W[k] \bsfU_W^*[k],
\end{equation}
which holds if and only if $\bsfU_W[k]$ diagonalizes $\bsfC[k]$, i.e., if the columns of $\bsfU_W[k]$ constitute an orthonormal basis for the subspace spanned by the $\Ns$ dominant eigenvectors of $\bsfC[k]$.

Now, we will attempt to maximize \eqref{equation:Witsenhausen_Uw} with respect to $\bsfU_F[k]$. Notice that the function to maximize is 
\begin{equation}
\begin{split}
{\cal R}(\bsfF[k],\bsfW[k]) &= \frac{1}{K}\sum_{k=0}^{K-1}\log_2\left(\prod_{i=1}^{\Ns}(1+\lambda_i(\bsfC[k]))\right) \\
	&\leq \frac{1}{K}\sum_{k=0}^{K-1}\log_2\left|\bI_{\Nr} + \frac{\SNR}{\Ns} \bsfH[k] \bsfU_F[k] \bm \Sigma_F^2[k] \bsfU_F^*[k] \bsfH^*[k] \right|.
\end{split}
	\label{equation:se_precoder}
\end{equation}
The main problem in \eqref{equation:se_precoder} comes from the characterization of the matrix $\bm \Sigma_F[k]$. 
Notice, however, that if we express $\sigma_\text{max}^2\left(\bsfF[k]\right) = \max\left\{\sigma_1^2\left(\bsfF[k]\right),\ldots,\sigma_{\Ns}^2\left(\bsfF[k]\right)\right\}$, the power constraints can be written as
\begin{equation}
	\frac{1}{\Ns}\sum_{k=0}^{K-1}\max\left\{\sigma_1^2\left(\bsfF[k]\right),\ldots,\sigma_{\Ns}^2\left(\bsfF[k]\right)\right\} \leq p_0.
	\label{equation:relaxed_constraints_2}
\end{equation}
Now, since the $\max\{\cdot\}$ function is convex \cite{Boyd_Convex}, and an affine transformation of convex functions is again convex, the constraints in \eqref{equation:relaxed_constraints_2} are also convex. Since the function in \eqref{equation:se_precoder} is log-concave, the optimization problem in \eqref{equation:opt_se_relaxed} is also convex (concave). Then, we know that a local maximum of that function is also the global maximizer of the function. The expression in \eqref{equation:relaxed_constraints_2} reveals that, given the maximum singular value of $\bsfF[k]$, the only constraint on the remaining $\Ns-1$ singular values 
is that they must be smaller than or equal to the maximum. Therefore, maximizing the functional in \eqref{equation:se_precoder} requires setting the different $\Ns$ singular values in $\bsfF[k]$ to the maximum for every subcarrier, which yields $\bm \Sigma_F^2[k] = \rho^2[k]\bI_{\Ns}$. This allows us to express \eqref{equation:se_precoder} as
\begin{equation}
	{\cal R}(\bsfF[k],\bsfW[k]) = \frac{1}{K}\sum_{k=0}^{K-1}\log_2\left|\bI_{\Ns} + \frac{\SNR}{\Ns} \rho^2[k] \bsfU_F[k] \bsfU_F^*[k] \bsfH^*[k] \bsfH[k]\right|.
\end{equation}
Now, we can apply Lemma 3 from \cite{Witsenhausen} such that
\begin{equation}
\begin{split}
	{\cal R}(\bsfF[k],\bsfW[k]) &\leq \frac{1}{K}\sum_{k=0}^{K-1}\log_2 \left( \prod_{i=1}^{\Nt}(1 + \frac{\rho^2[k] \SNR}{\Ns} \lambda_i\left(\bsfU_F[k] \bsfU_F^*[k]\right)\lambda_i\left(\bsfH^*[k] \bsfH[k]\right))\right) \\
	&= \frac{1}{K} \sum_{k=0}^{K-1}\log_2\left(\prod_{i=1}^{\Ns}(1 + \frac{\rho^2[k]\SNR}{\Ns} \lambda_i\left(\bsfU_F[k] \bsfU_F^*[k]\right) \lambda_i\left(\bsfH^*[k] \bsfH[k] \right))\right) \\
	&= \frac{1}{K} \sum_{k=0}^{K-1}\log_2\left|\bI_{\Ns} + \frac{\rho^2[k] \SNR}{\Ns} \tilde{\bm \Sigma}_H^2[k] \right|.
\end{split}
\label{equation:Witsenhausen}
\end{equation}
Equality in \eqref{equation:Witsenhausen} is achieved if $\bsfU_F[k] \bsfU_F^*[k]$ and $\bsfH^*[k] \bsfH[k]$ commute. If two normal matrices commute, they are simultaneously diagonalizable. The reverse also holds true. Then, the column vectors in $\bsfU_F[k]$ that achieve equality in \eqref{equation:Witsenhausen} are the $\Ns$ eigenvectors of $\bsfH^*[k] \bsfH[k] = \bsfU_H[k] \bm \Sigma_H^2[k] \bsfU_H^*[k]$ corresponding to the $\Ns$ largest eigenvalues contained in $\bm \Sigma_H^2[k]$. 
Finally, the function to optimize is given by
\begin{equation}
\begin{split}
	{\cal R}(\bsfF[k],\bsfW[k]) &= \frac{1}{K}\sum_{k=0}^{K-1}\log_2\left|\bI_{\Ns} + \frac{\rho^2[k] \SNR}{\Ns} \bm \Sigma_H^2[k] \right|, \\
	&\text{subject to} \qquad \frac{1}{\Ns}\sum_{k=0}^{K-1}\rho^2[k] \leq p_0.
\end{split}
\label{equation:opt_se_relaxed_proof}
\end{equation}
The optimization problem in \eqref{equation:opt_se_relaxed_proof} is convex, and thus can be solved using any convex optimization tool. This concludes the proof.
\end{proof}

Notice that the optimum precoder for problem \eqref{equation:opt_se_relaxed} is a scaled semiunitary matrix with uniform power allocation across the $\Ns$ different data streams forwarded through every subcarrier $0\leq k \leq K-1$. If the precoder $\bsfU_F[k]$ from Theorem 1 is the optimum solution to the relaxed problem, it is also feasible for the original problem in \eqref{equation:opt_se_original}, although suboptimum. Then, the performance of the precoder $\bsfF[k]$ given by Theorem 1 can be improved if we define another precoder $\bsfF[k] = \bsfU_F[k] \bm \Sigma_F[k]$, with $\bm \Sigma_F[k] = \diag\{\rho_1[k],\ldots,\rho_{\Ns}[k]\}$ and optimize the power allocation coefficients $\rho_i[k]$, $1\leq i \leq \Ns$ for the different subcarriers, subject to the per-antenna power constraints. In fact, such choice of frequency-selective precoder leads to an upper bound for the spectral efficiency maximization problem in \eqref{equation:opt_se_original}. This can be shown using Sylvester's theorem and Lemma 3 from \cite{Witsenhausen} in \eqref{equation:se_precoder} as
\begin{equation}
\begin{split}
	{\cal R}(\bsfF[k],\bsfW[k]) &\leq \frac{1}{K}\sum_{k=0}^{K-1}\log_2\left|\bI_{\Nr} + \frac{\SNR}{\Ns}\bsfH[k]\bsfU_F[k]\bm \Sigma_F^2[k] \bsfU_F^*[k] \bsfH^*[k]\right| \\
	&= \frac{1}{K}\sum_{k=0}^{K-1}\log_2\left|\bI_{\Nt} + \frac{\SNR}{\Ns}\bsfU_F[k]\bm \Sigma_F^2[k] \bsfU_F^*[k] \bsfH^*[k] \bsfH[k] \right| \\
	&\leq \frac{1}{K}\sum_{k=0}^{K-1}\log_2\left(\prod_{i=1}^{\Nt}\left(1 + \frac{\SNR}{\Ns}\lambda_i\left(\bsfU_F[k]\bm \Sigma_F^2[k] \bsfU_F^*[k]\right)\lambda\left(\bsfH^*[k]\bsfH[k]\right) \right) \right) \\
	&= \frac{1}{K}\sum_{k=0}^{K-1}\log_2\left(\prod_{i=1}^{\Ns}\left(1 + \frac{\SNR}{\Ns}\lambda_i\left(\bm \Sigma_F^2[k]\right) \lambda_i\left(\bm \Sigma_H^2[k]\right)\right)\right) \\
	&= \frac{1}{K}\sum_{k=0}^{K-1}\log_2\left|\bI_{\Ns} + \frac{\SNR}{\Ns} \bm \Sigma_F^2[k] \tilde{\bm \Sigma}_H^2[k]\right|,
\end{split}
\label{equation:upper_bound}
\end{equation}
where $\tilde{\bm \Sigma}_H^2[k] \in \mathbb{R}^{\Ns \times \Ns}$ was already defined as the diagonal matrix containing the $\Ns$ dominant singular values of $\bm \Sigma_H[k]$. Notice that the per-antenna power constraints impose conditions on both the singular values and the left singular vectors of the unconstrained precoder, such that both $\bm \Sigma_F[k]$ and $\bsfU_F[k]$ are coupled and need to be optimized simultaneously. If, however, a certain $\bsfU_F[k]$ is feasible for the relaxed problem in \eqref{equation:opt_se_relaxed}, then it is also feasible for the original problem in \eqref{equation:opt_se_original}. Thus, the upper bound for the spectral efficiency in \eqref{equation:upper_bound} can be attained without violating the power constraints by taking $\bsfU_F[k]$ as in Theorem 1 and optimizing the power allocation coefficients in the diagonal of $\bm \Sigma_F[k]$. 

In view of this, the spectral efficiency maximization problem can be stated as
\begin{equation}
\begin{split}
	& \underset{\underset{0\leq l \leq K-1}{\bm \Sigma_F[l]}}{\max}\,\frac{1}{K}\sum_{k=0}^{K-1}\log_2\left|\bI_{\Ns} + \frac{\SNR}{\Ns} \bm \Sigma_F^2[k] \tilde{\bm \Sigma}_H^2[k]\right| \\
	&\text{subject to} \qquad \frac{1}{\Ns}\bee_j^*\left(\sum_{k=0}^{K-1}\bsfU_F[k] \bm \Sigma_F^2[k] \bsfU_F^*[k]\right)\bee_j \leq p_j, \, j = 1,\ldots,\Nt.
\end{split}
\label{equation:se_optimized}
\end{equation}
If we denote the $(j,\ell)$-th element of $\bsfU_F[k]$ by $u_{j\ell}[k]$, $1\leq j \leq \Nt$, $1\leq \ell \leq \Ns$, then the problem in \eqref{equation:se_optimized} can be expressed in scalar form as
\begin{equation}
\begin{split}
	&\underset{\underset{0\leq l \leq K-1}{\rho_i[l]}}{\max}\,\frac{1}{K}\sum_{k=0}^{K-1}\sum_{\ell=1}^{\Ns}\log_2\left(1 + \frac{\SNR}{\Ns} \sigma_{H,\ell}^2[k] \rho_\ell^2[k]\right) \\
	&\text{subject to} \qquad 0 \leq \frac{1}{\Ns}\sum_{k=0}^{K-1}\sum_{\ell=1}^{\Ns}|u_{j\ell}[k]|^2 \rho_\ell^2[k] \leq p_j, \, j = 1,\ldots,\Nt.
\end{split}
\label{equation:se_optimized_explicit}
\end{equation}
Since the problem in \eqref{equation:se_optimized_explicit} is convex, the power allocation coefficients $\{\rho_\ell[k]\}$ can be efficiently found using any convex optimization tool. The result is a modified space-frequency waterfilling algorithm, in which the \text{effective $\SNR$} per subcarrier determines how the per-antenna power budget is splitted into the different subchannels.

\section{Design of frequency-selective hybrid precoders}
\label{sec:hybrid_precoders}

In this section, we propose a design method for the frequency-selective hybrid precoders with per-antenna power constraints. Let us $\bsfFrf \in \mathbb{C}^{\Nt \times \Lt}$ to denote the frequency-flat analog precoder, and $\bsfFbb[k] \in \mathbb{C}^{\Lt \times \Ns}$ to denote the frequency-selective digital precoder. In \cite{AyaRajAbuPiHea:Spatially-Sparse-Precoding:14} it was shown that, under certain approximations, maximizing the mutual information achieved by a hybrid precoder $\bF = \bF_\text{RF} \bF_\text{BB}$ over a narrowband \ac{mmWave} \ac{MIMO} channel $\bH$ was equivalent to minimizing the chordal distance between the optimum unconstrained precoder given by the principal $\Ns$ eigenvectors of $\bH^*\bH$ and the hybrid precoder $\bF$, subject to the hardware constraints. Furthermore, we know from Theorem 1 that any orthonormal basis of the subspace spanned by the columns in $\tilde{\bV}_H[k]$ is optimal for problem \eqref{equation:opt_se_relaxed}. These facts motivate the design of the frequency-selective hybrid precoders as
\begin{eqnarray}
\lefteqn{\!\!\!\!\!\!\!\!\! 
	\underset{\bsfFrf,\{\bsfFbb[k]\}_{k=0}^{K-1}}{\max\,}\sum_{k=0}^{K-1}\|\bsfF^*[k]\bsfFrf\bsfFbb[k]\|_F^2}, 	\label{equation:opt_problem_precoders} \\
		\mbox{subject to} && \left\{\begin{array}{cc} \bsfFrf \in {\cal M}^{\Nt \times \Lt}(Q_\text{t}), & \\
		 \frac{1}{\Ns}\bee_j^*\left(\sum_{k=0}^{K-1}\bsfFrf\bsfFbb[k]\bsfFbbher[k]\bsfFrfher\right)\bee_j \leq p_j, & j = 1,\ldots,\Nt, \\ \end{array} \right. \nonumber
\end{eqnarray}
where $\bsfF[k] \in \mathbb{C}^{\Nt \times \Ns}$ is the frequency-selective unconstrained precoder found in Section \ref{section:all-digital}. Since finding the solution to problem \eqref{equation:opt_problem_precoders} is intractable due to the hardware constraints, we will find a suboptimal approximation for the precoding matrices $\bsfFrf$ and $\bsfFbb[k]$. 

The problem in \eqref{equation:opt_problem_precoders} requires finding a suitable frequency-flat matrix $\bsfFrf \in \mathcal{M}^{\Nt \times \Lt}$ that spans the subspace spanned by the different $\{\bsfF[k]\}_{k=0}^{K-1}$. In the following, we describe a method to design such RF precoder.

In \cite{KiranNuriaRobert} it was proven that, if the frequency-selective channel $\bsfH[k]$ is such that $\sum_{c=1}^{C}R_c \leq \min(\Nt,\Nr)$, then there exists a semi-unitary frequency-flat matrix $\bsfF_\text{ff} \in \mathbb{C}^{\Nt \times \Ns}$ such that the principal $\Ns$ left singular vectors can be expressed as $\tilde{\bsfV}_H[k] = \bsfF_\text{ff} \bsfQ[k]$, with $\bsfQ[k] \in \mathbb{C}^{\Ns \times \Ns}$ a unitary matrix. In \ac{mmWave} frequency-selective channels, the condition $\sum_{c=1}^{C}R_c \leq \min(\Nt,\Nr)$ does not generally hold. Despite this, we can still make an approximation to find suitable hybrid precoders and combiners as follows. Let us define the semi-unitary matrices $\bsfP \in \mathbb{C}^{\Nt \times \Lt}$ and $\bsfQ_P[k] \in \mathbb{C}^{\Lt \times \Ns}$. Then, we can approximate the optimum precoder as $\bsfF[k] \approx \bsfP \bsfQ_P[k]$. The matrix $\bsfP$ can be found as the solution to the following problem:
\begin{equation}
	\underset{\bsfP}{\arg\,\max\,}\sum_{k=0}^{K-1}\|\bsfF^*[k]\bsfP \bsfQ_P[k]\|_F^2.
	\label{equation:opt_freq_flat_low_rank}
\end{equation}
Let us define a matrix $\bsfZ[k] \in \mathbb{C}^{\Ns \times \Lt}$, $\bsfZ[k] = \bsfF^*[k] \bsfP$, with SVD $\bsfZ[k] = \bsfU_Z[k] \bm \Sigma_Z[k] \bsfV_Z^*[k]$. Since $\bsfF[k] \approx \bsfP \bsfQ_P[k]$, we can use Von Neumann's trace inequality in \eqref{equation:opt_freq_flat_low_rank}, such that
\begin{equation}
\begin{split}
	\sum_{k=0}^{K-1}\|\bsfF^*[k] \bsfP \bsfQ_P[k]\|_F^2 &= \sum_{k=0}^{K-1}\trace\{\bsfF^*[k] \bsfP \bsfQ_P[k] \bsfQ_P^*[k] \bsfP^* \bsfF[k]\} \\
	&\overset{(a)}{\leq} \sum_{k=0}^{K-1}\trace\{\bm \Sigma_Z^2[k]\},
\end{split}
\label{equation:opt_freq_flat_Von_Neumann}
\end{equation}
where $(a)$ follows from the fact that $\bsfQ_P[k] \bsfQ_P^*[k]$ has eigenvalues $\lambda_i(\bsfQ_P[k] \bsfQ_P^*[k]) = 1$, $1 \leq i \leq \Ns$. The upper bound in \eqref{equation:opt_freq_flat_Von_Neumann} is achieved under the approximation $\bsfF[k] \approx \bsfP \bsfQ_P[k]$ since $\bsfF^*[k]\bsfP \approx \bsfQ_P^*[k] \bsfP^* \bsfP \approx \bsfQ_P^*[k]$, which of course spans the subspace spanned by $\bsfQ_P[k]$. Then, the upper bound in \eqref{equation:opt_freq_flat_Von_Neumann} is maximized if 
\begin{equation}
	\sum_{k=0}^{K-1}\trace\{\bm \Sigma_Z^2[k]\} = \sum_{k=0}^{K-1}\|\bsfF^*[k] \bsfP\|_F^2
	\label{equation:max_upper_bound_freq_flat}
\end{equation}
is maximum. The cost function in the right-hand side of \eqref{equation:max_upper_bound_freq_flat} can be developed as
\begin{equation}
\begin{split}
	\sum_{k=0}^{K-1}\|\bsfF^*[k]\bsfP\|_F^2 &= \sum_{k=0}^{K-1}\trace\left\{\bsfP^*\bsfF[k] \bsfF^*[k] \bsfP\right\} \\
	&= \trace\left\{\bsfP^*\underbrace{\left(\sum_{k=0}^{K-1}\bsfF[k]\bsfF^*[k]\right)}_{\bsfT}\bsfP\right\} \\
	&\leq \trace\left\{\left[\bm \Lambda_T\right]_{1:\Lt,1:\Lt}\right\},
\end{split}
\label{equation:subopt_RF_precoder}
\end{equation}
Thus, according to \eqref{equation:subopt_RF_precoder}, the frequency-flat matrix that approximately maximizes \eqref{equation:opt_freq_flat_low_rank} is given by the first eigenvectors of $\bsfT = \bsfU_T \bm \Lambda_T \bsfU_T^*$. Although not explicitly derived, this solution has already been adopted in \cite{SPAWC_2018} to initialize the hybrid precoding and combining design algorithm. Therefore, it is reasonable to design the RF precoder $\bsfFrf$ as a function of this matrix. Let us denote the $(j,k)$-th element in $\bsfU_T$ as $\sfu_{j,k}$. Then, a sensible closed-form expression for $(\bsfFrf)_{j,k}$ is 
\begin{equation}
	(\bsfFrf)_{j,k} = e^{jQ_\text{t}(\sfu_{j,k})}, \left\{\begin{array}{cc} 
	j = 1,\ldots,\Nt, & \\
	k = 1,\ldots,\Lt, & \\ \end{array}\right.
	\label{equation:RF_precoder}
\end{equation}
where $Q_\text{t}(\sfu_{j,k})$ models the mitread phase quantization process according to the number of quantization bits for the transmit phase-shifters in the hybrid architecture.

Besides being driven by an approximation, the solution in \eqref{equation:RF_precoder} assumes that the frequency-flat matrix $\bsfP$ is semi-unitary, which indeed adds further constraints to the design of the RF precoder $\bsfFrf$. That solution, however, can be proven to maximize the lower bound of the function in \eqref{equation:opt_problem_precoders} when the per-antenna power constraints are ignored. This is characterized by the following result:

\textbf{Lemma 2}. Let us consider an SVD of the baseband precoder $\bsfFbb[k] = \bsfU_{F,\text{BB}}[k] \bm \Sigma_{F,\text{BB}}[k] \bsfV_{F,\text{BB}}^*[k]$, with $\bsfU_{F,\text{BB}}[k] \in \mathbb{C}^{\Lt \times \Ns}$, $\bm \Sigma_{F,\text{BB}}[k] \in \mathbb{C}^{\Ns \times \Ns}$, and $\bsfV_{F,\text{BB}}[k] \in \mathbb{C}^{\Ns \times \Ns}$. If the influence of the per-antenna power constraints on the singular vectors of $\bsfFrf$ and $\bsfFbb[k]$ is neglected, a lower bound for the cost function in \eqref{equation:opt_problem_precoders} is given by
\begin{equation}
	\sum_{k=0}^{K-1}\|\bsfF^*[k] \bsfFrf \bsfFbb[k]\|_F^2 \geq \left(\underset{0\leq k \leq K-1}{\min}\frac{1}{\|\bm \Sigma_{F,\text{BB}}^{-2}[k]\|}\right)\sum_{k=0}^{K-1}\|\bsfF^*[k] \bsfFrf\|_F^2,
	\label{equation:lower_bound}
\end{equation}
where $\|\bA\|$ denotes the spectral norm of a matrix $\bA$.

\textit{Proof}. Let us define $\bsfG[k] = \bsfF^*[k]\bsfFrf$, with an SVD $\bsfG[k] = \bsfU_G[k] \bm \Sigma_G[k] \bsfV_G^*[k]$, with $\bsfU_G[k] \in \mathbb{C}^{\Ns \times \Ns}$, $\bm \Sigma_G[k] \in \mathbb{C}^{\Ns \times \Ns}$, and $\bsfV_G[k] \in \mathbb{C}^{\Lt \times \Ns}$. Now, consider the cost function in \eqref{equation:opt_problem_precoders}, which can be developed as
\begin{equation}
\begin{split}
	\sum_{k=0}^{K-1}\|\bsfF^*[k] \bsfFrf \bsfFbb[k]\|_F^2 &= \sum_{k=0}^{K-1}\trace\{\bsfF^*[k] \bsfFrf \bsfFbb[k] \bsfFbbher[k] \bsfFrfher \bsfF[k]\} \\
	&\overset{(a)}{\leq} \sum_{k=0}^{K-1}\trace\{\bm \Sigma_G^2[k] \bm \Sigma_{F,\text{BB}}^2[k]\} \\
	&\geq \sum_{k=0}^{K-1}\sigma_\text{min}\left(\bm \Sigma_{F,\text{BB}}^2[k]\right)\trace\{\bm \Sigma_G^2[k]\} \\
	&=\sum_{k=0}^{K-1}\|\bsfF^*[k] \bsfFrf\|_F^2 \|\bm \Sigma_{F,\text{BB}}^{-2}[k]\|^{-1} \\
	&\geq\left(\underset{0\leq k \leq K-1}{\min}\frac{1}{\|\bm \Sigma_{F,\text{BB}}^{-2}[k]\|}\right) \sum_{k=0}^{K-1}\|\bsfF^*[k] \bsfFrf\|_F^2,
\end{split}
\label{equation:lower_bound_RF_precoder}
\end{equation}
where $(a)$ follows from Von Neumann's trace inequality. Therefore, minimizing the chordal distance between the unconstrained optimum precoder and the analog precoder maximizes the lower bound of the original cost function in \eqref{equation:opt_problem_precoders}. This completes the proof.

Now, with the RF precoder in \eqref{equation:RF_precoder}, the baseband precoders $\bsfFbb[k]$, $0\leq k \leq K-1$, can be now found by solving \eqref{equation:opt_problem_precoders} only as a function of $\bsfFbb[k]$:
\begin{eqnarray}
\lefteqn{\!\!\!\!\!\!\!\!\! 
	\underset{\{\bsfFbb[k]\}_{k=0}^{K-1}}{\max\,}\sum_{k=0}^{K-1}\|\bsfF^*[k]\bsfFrf\bsfFbb[k]\|_F^2}, 	\label{equation:opt_problem_baseband_precoder} \\
	\mbox{subject to} && \frac{1}{\Ns}\bee_j^*\left(\sum_{k=0}^{K-1}\bsfFrf\bsfFbb[k]\bsfFbbher[k]\bsfFrfher\right) \bee_j \leq p_j, \quad j = 1,\ldots,\Nt. \nonumber		
\end{eqnarray}
Let us consider again the SVD $\bsfFbb[k] = \bsfU_F[k] \bm \Sigma_F[k] \bsfV_F^*[k]$, with $\bsfU_F[k] \in \mathbb{C}^{\Lt \times \Ns}$, $\bm \Sigma_F[k] \in \mathbb{C}^{\Ns \times \Ns}$, and $\bsfV_F[k] \in \mathbb{C}^{\Ns \times \Ns}$. Let us denote $\bsfG[k] = \bsfF^*[k] \bsfFrf$, with SVD $\bsfG[k] = \bsfU_G[k] \bm \Sigma_G[k] \bsfV_G^*[k]$, with $\bsfU_G[k] \in \mathbb{C}^{\Ns \times \Ns}$, $\bm \Sigma_G[k] \in \mathbb{C}^{\Ns \times \Ns}$, $\bV_G[k] \in \mathbb{C}^{\Ns \times \Lt}$. Notice that the Frobenius norm is unitarily invariant, such that the objective function in \eqref{equation:opt_problem_baseband_precoder} becomes
\begin{equation}
\begin{split}
	\sum_{k=0}^{K-1}\|\bsfG[k] \bsfFbb[k]\|_F^2 &= \sum_{k=0}^{K-1}\trace\left\{\bsfFbb^*[k] \bsfG^*[k] \bsfG[k] \bsfFbb[k]\right\} \\
	&= \sum_{k=0}^{K-1}\trace\left\{\bsfV_F[k] \bm \Sigma_F[k] \bsfU_F^*[k] \bsfG^*[k] \bsfG[k] \bsfU_F[k] \bm \Sigma_F[k] \bsfV_F^*[k] \right\} \\
	&= \sum_{k=0}^{K-1}\trace\left\{\bsfG[k] \bsfU_F[k] \bm \Sigma_F^2[k] \bsfU_F^*[k] \bsfG^*[k]\right\}.
\end{split}
\label{equation:cost_function_baseband_precoder}
\end{equation}
In view of \eqref{equation:cost_function_baseband_precoder}, and since the constraints in \eqref{equation:opt_problem_baseband_precoder} do not depend on $\bsfV_F[k]$, we can set $\bsfV_F[k] = \bI_{\Ns}$ without loss of generality. Finding the solution to problem \eqref{equation:opt_problem_baseband_precoder} requires designing both $\bsfU_F[k]$ and $\bm \Sigma_F[k]$ taking into account the per-antenna power constraints. Since a closed-form solution to this problem is difficult to characterize, we propose to find an approximate suboptimal solution as follows.
 The singular values in $\bm \Sigma_F[k]$ and $\bm \Sigma_G[k]$ are assumed sorted in descending order. Then, using Von Neumann's trace inequality \cite{VonNeumann}, the cost function in \eqref{equation:cost_function_baseband_precoder} can be upper bounded as
\begin{equation}
\begin{split}
	\sum_{k=0}^{K-1}\|\bsfG[k] \bsfFbb[k]\|_F^2 &\leq \sum_{k=0}^{K-1}\trace\left\{\bm \Sigma_G^2[k] \bm \Sigma_F^2[k]\right\}.
\end{split}
\label{equation:suboptimal_U_F}
\end{equation}
Equality in \eqref{equation:suboptimal_U_F} is achieved if $\bsfU_F[k] = \bsfV_G^*[k]$, similarly to Lemma 2. Notice, however, that the upper bound in \eqref{equation:suboptimal_U_F} is achieved if we neglect the influence of per-antenna power constraints on $\bsfU_F[k]$ for the different subcarriers. Thereby, setting $\bsfU_F[k] = \bsfV_G^*[k]$ is an approximate solution to problem \eqref{equation:opt_problem_baseband_precoder}. Thus, we propose to choose the baseband precoder as $\bsfFbb[k] = \bsfV_G^*[k] \bm \Sigma_F[k]$. In \cite{Globecom_16_per_antenna}, it was chosen to maximize the upper bound in \eqref{equation:suboptimal_U_F} as a function of $\bm \Sigma_F[k]$ to find this power allocation matrix. This is a good approximation to design $\bm \Sigma_F$ in the narrowband scenario (i.e. $K = 1$), but it is not a suitable strategy for frequency-selective channels. To illustrate this, let us consider the maximization of the upper bound in \eqref{equation:suboptimal_U_F} as a function of the per-antenna power constraints:
\begin{eqnarray}
\lefteqn{\!\!\!\!\!\!\!\!\! 
	\underset{\{\bm \Sigma_F[k]\}_{k=0}^{K-1}}{\max\,}\sum_{k=0}^{K-1}\trace\left\{\bm \Sigma_G^2[k] \bm \Sigma_F^2[k]\right\}}, 	\label{equation:opt_max_trace} \\
		\mbox{subject to} && \left\{\begin{array}{cc} \frac{1}{\Ns}\bee_j^*\left(\sum_{k=0}^{K-1}\bsfFrf\bsfV_G^*[k]\bm \Sigma_F^2[k] \bsfV_G[k]\bsfFrfher\right)\bee_j \leq p_j, \quad j = 1,\ldots,\Nt, \\
		\bm \Sigma_F^2[k] \succeq \bm 0. \\ \end{array}\right. \nonumber
\end{eqnarray}
The objective function in \eqref{equation:opt_max_trace} is an affine function of $\Ns$ product terms. The constraints in \eqref{equation:opt_max_trace} do not bound each individual entry in $\bm \Sigma_F^2[k]$, but there is only a constraint affecting the sum of contributions coming from the different $\bm \Sigma_F^2[k]$, $k = 0,\ldots,K-1$. If $K = 1$, the first set of $\Nt$ constraints in \eqref{equation:opt_max_trace} implicitly upper bounds each entry in $\bm \Sigma_F^2[k]$. For $K > 1$, optimizing \eqref{equation:opt_max_trace} will intuitively result in the power budget $p_j$, for $1\leq j \leq \Nt$ being splitted into those entries $\left[\bm \Sigma_F^2[k]\right]_{i,i}$ that correspond to the largest $\left[\bm \Sigma_G^2[k]\right]_{i,i}$, such that the per-antenna constraints are not violated. This, however, does not ensure that the spectral efficiency is maximized. For this reason, we choose not to maximize the upper bound in \eqref{equation:opt_max_trace}, and maximize the spectral efficiency, instead.


Remark: $\bsfU_F[k] = \bsfV_G^*[k]$ is not necessarily the optimal solution unless the per-antenna power constraints are ignored. In view of \eqref{equation:opt_problem_baseband_precoder}, the optimization variables $\bsfU_F[k]$ and $\bm \Sigma_F[k]$ are coupled. Thus, this particular choice of $\bsfU_F[k]$ sets further constraints on the design of $\bm \Sigma_F[k]$ in \eqref{equation:opt_problem_baseband_precoder}. However, as we will show in numerical results, this practical choice allows us to obtain values of spectral efficiency close to the perfect CSI design with a total power constraint.

\section{Design of frequency-selective hybrid combiners}
\label{sec:hybrid_combiners}

In this section, we propose a design method for the frequency-selective hybrid combiners. To start the discussion, let us consider the all-digital solution given by Theorem 1:
\begin{equation}
	\bsfW[k] = \tilde{\bsfU}_H[k] \bsfR_W[k].
	\label{equation:optimal_Th1}
\end{equation}
In view of \eqref{equation:optimal_Th1}, it is clear that any basis of the $\Ns$-dimensional subspace spanned by the columns of $\tilde{\bsfU}_H[k]$ is optimal for problem \eqref{equation:opt_se_relaxed}. Furthermore, it is well-known that there is no loss of optimality in terms of spectral efficiency if the hybrid combiner $\bsfWrf \bsfWbb[k]$ has orthonormal columns. These facts motivate the following design for the hybrid combiner:
\begin{eqnarray}
\lefteqn{\!\!\!\!\!\!\!\!\! 
	\underset{\bsfWrf,\{\bsfWbb[k]\}_{k=0}^{K-1}}{\max\,}\sum_{k=0}^{K-1} \|\tilde{\bsfU}_H^*[k] \bsfWrf \bsfWbb[k]\|_F^2}	\label{equation:opt_hybrid_combiners} \\
		\mbox{subject to} && \left\{\begin{array}{cc} 
		\bsfWrf \in {\cal M}^{\Nr \times \Lr}(Q_\text{r}), & \\
		\left(\bsfWrf \bsfWbb[k]\right)^*\left(\bsfWrf \bsfWbb[k]\right) = \bI_{\Ns}. \end{array}\right. \nonumber
\end{eqnarray} 
To find a solution to problem \eqref{equation:opt_hybrid_combiners}, we propose to use a similar approach to that in Section \ref{sec:hybrid_precoders}. We will first obtain a reasonable approximation for the RF combiner $\bsfWrf$, and then optimize the baseband precoders $\bsfWbb[k]$ for the different subcarriers. Let us consider an SVD of the baseband combiners $\bsfWbb[k] = \bsfU_{W,\text{BB}}[k] \bm \Sigma_{W,\text{BB}}[k] \bsfV_{W,\text{BB}}^*[k]$, with $\bsfU_{W,\text{BB}}[k] \in \mathbb{C}^{\Lr \times \Ns}$, $\bm \Sigma_{W,\text{BB}}[k]$, and $\bsfV_{W,\text{BB}}[k] \in \mathbb{C}^{\Ns \times \Ns}$. Then, similarly to Section \ref{sec:hybrid_precoders}, a lower bound for \eqref{equation:opt_hybrid_combiners} is given by:
\begin{equation}
	\sum_{k=0}^{K-1}\|\tilde{\bsfU}_H^*[k]\bsfWrf \bsfWbb[k]\|_F^2 \geq \left(\underset{0\leq k \leq K-1}{\min\,}\frac{1}{\|\bm \Sigma_{W,\text{BB}}^{-2}[k]\|}\right) \sum_{k=0}^{K-1}\|\tilde{\bsfU}_H^*[k] \bsfWrf\|_F^2.
	\label{equation:relaxed_opt_combiner}
\end{equation}
Then, the problem of finding the RF combiner can be stated as
\begin{eqnarray*}
	\underset{\bsfWrf}{\max\,}\sum_{k=0}^{K-1}\|\tilde{\bsfU}_H^*[k] \bsfWrf\|_F^2 \label{equation:problem_RF_combiner}, \\
	\mbox{subject to} \quad \|\bsfWrf\|_F^2 = \Lr \Nr, \nonumber
\end{eqnarray*}
where the constraint in \eqref{equation:problem_RF_combiner} is a convex relaxation of the constraint $|(\bsfWrf)_{j,k}| = 1$, $j = 1,\ldots,\Nr$, $k = 1,\ldots,\Lr$.
If we develop the cost function in \eqref{equation:problem_RF_combiner}, we obtain
\begin{equation}
\begin{split}
	\sum_{k=0}^{K-1}\|\tilde{\bsfU}_H^*[k]\bsfWrf\|_F^2 &= \sum_{k=0}^{K-1}\trace\left\{\tilde{\bsfU}_H^*[k] \bsfWrf \bsfWrf^* \tilde{\bsfU}_H[k]\right\} \\
	&= \trace\left\{\bsfWrf^* \underbrace{\left( \sum_{k=0}^{K-1} \tilde{\bsfU}_H[k] \tilde{\bsfU}_H^*[k]\right)}_{\bsfS} \bsfWrf \right\} \\
	&\leq \trace\left\{\left[\bm \Lambda_S\right]_{1:\Lr, 1:\Lr}\right\},
\end{split}
\label{equation:RF_combiner}
\end{equation}
where the eigendecomposition $\bsfS = \bsfU_S \bm \Lambda_S \bsfU_S^*$ has been used, and the upper bound in \eqref{equation:RF_combiner} follows from using Von Neumann's trace inequality. Such upper bound is achieved by setting $\bsfW_{\text{RF}}$ to the first $\Lr$ eigenvectors of $\bsfS$, which are henceforth denoted by $\tilde{\bsfU}_S$, and $\bsfV_{W,\text{RF}}$ can be set to any matrix having orthonormal columns. Therefore, it is sensible to set the RF combiner $\bsfWrf$ as
\begin{equation}
	(\bsfWrf)_{j,k} = e^{jQ_\text{r}((\tilde{\bsfU}_S)_{j,k})}, \left\{\begin{array}{cc} 
	j = 1,\ldots,\Nr, & \\
	k = 1,\ldots,\Lr,& \\ \end{array}\right.
\end{equation}
with SVD $\bsfWrf = \bsfU_{W,\text{RF}} \bm \Sigma_{W,\text{RF}} \bsfV_{W,\text{RF}}^*$, $\bsfU_{W,\text{RF}} \in \mathbb{C}^{\Nt \times \Lr}$, $\bm \Sigma_{W,\text{RF}} \in \mathbb{C}^{\Lr \times \Lr}$, and $\bsfV_{W,\text{RF}} \in \mathbb{C}^{\Lr \times \Lr}$.

Now, consider the set of matrices $\bsfWbb[k]$ satisfying the second constraint in \eqref{equation:opt_hybrid_combiners}. Such constraint can be expressed as
\begin{equation}
\begin{split}
	\bsfWbbher[k] \bsfWrfher \bsfWrf \bsfWbb[k] &= \bI_{\Ns} \\
	\bsfWbb^*[k] \bsfV_{W,\text{RF}}^* \bm \Sigma_{W,\text{RF}}^2 \bsfV_{W,\text{RF}} \bsfWbb[k] &= \bI_{\Ns}.
\end{split}
\label{equation:constraint_semiunitary_combiner}
\end{equation}
The constraint in \eqref{equation:constraint_semiunitary_combiner} can be noticed to be fulfilled by the baseband combiners $\bsfWbb[k] = \bsfU_{W,\text{RF}} \bm \Sigma_{W,\text{RF}}^{-1} \bsfZ_W[k]$, where $\bsfZ_W[k] \in \mathbb{C}^{\Lr \times \Ns}$ is a matrix having orthonormal columns. Finally, the problem of optimizing $\bsfZ_W[k]$ can be stated as
\begin{eqnarray}
	\underset{\{\bsfZ_W[k]\}_{k=0}^{K-1}}{\max\,}\sum_{k=0}^{K-1}\left\|\tilde{\bsfU}_H^*[k] \bsfWrf \bsfZ_W[k]\right\|_F^2 \label{equation:baseband_combiner_Zw} \\
	\mbox{subject to}\quad \bsfZ_W^*[k] \bsfZ_W[k] = \bI_{\Ns}. \nonumber
\end{eqnarray}
Since there is no constraint on the different $\bsfZ_W[k]$ beyond being semi-unitary, each of the terms in \eqref{equation:baseband_combiner_Zw} can be individually optimized for each subcarrier. Every term is maximized when $\bsfZ_W[k]$ spans the same subspace as the $\Ns$ left singular vectors of $\bsfY[k] = \bsfWrfher \tilde{\bsfU}_H[k]$. If we consider a reduced SVD $\bsfY[k] = \bsfU_Y[k] \bm \Sigma_Y[k] \bsfV_Y^*[k]$, with $\bsfU_Y[k] \in \mathbb{C}^{\Lt \times \Ns}$, $\bm \Sigma_Y[k] \in \mathbb{C}^{\Ns \times \Ns}$, and $\bsfV_Y[k] \in \mathbb{C}^{\Ns \times \Ns}$, then the choice $\bsfZ_W[k] = \bsfV_Y[k]$ ensures that \eqref{equation:baseband_combiner_Zw} is maximized.

Finally, the power allocation matrix $\bm \Sigma_F[k]$ in Section \ref{sec:hybrid_precoders} can be found by maximizing the spectral efficiency in \eqref{equation:spectral_efficiency}, taking into account the designed hybrid precoders (except for $\bm \Sigma_F[k]$) and combiners. Let $\bsfH_\text{eff}[k] \in \mathbb{C}^{\Ns \times \Ns}$ be defined as
\begin{equation}
	\bsfH_\text{eff}[k] = \bsfV_Y^*[k] \bm \Sigma_{W,\text{RF}}^{-1} \bsfU_{W,\text{RF}}^* \bsfWrfher \bsfH[k] \bsfFrf \bsfV_G^*[k], \quad k = 0,\ldots,K-1.
\end{equation}
Then, the spectral efficiency achieved by the hybrid precoders and combiners is given by
\begin{eqnarray}
\lefteqn{\!\!\!\!\!\!\!\!\! 
	\underset{\{\bm \Sigma_F[k]\}_{k=0}^{K-1}}{\max\,}\frac{1}{K}\sum_{k=0}^{K-1}\log_2\left|\bI_{\Ns} + \bsfH_\text{eff}[k] \bm \Sigma_F^2[k] \bsfH_\text{eff}^*[k]\right|}, 	\label{equation:opt_max_se} \\
		\mbox{subject to} && \left\{\begin{array}{cc} \frac{1}{\Ns}\bee_j^*\left(\sum_{k=0}^{K-1}\bsfFrf\bsfV_G^*[k]\bm \Sigma_F^2[k] \bsfV_G[k]\bsfFrfher\right)\bee_j \leq p_j, \quad j = 1,\ldots,\Nt, \\
		\bm \Sigma_F^2[k] \succeq \bm 0, \\ \end{array}\right. \nonumber
\end{eqnarray}
which involves the maximization of an affine function of log-concave functions, subject to linear constraints on the entries of $\bm \Sigma_F^2[k]$. Therefore, this problem can be efficiently solved using any convex optimization tool.

\section{Numerical Results}

In this section, the main numerical results obtained with the proposed precoding and combining algorithms are presented. To obtain these results, we perform Monte Carlo simulations averaged over $100$ trials to evaluate the ergodic spectral efficiency as a function of different system parameters. Owing to lack or prior work on hybrid precoding and combining under frequency-selective \ac{mmWave} \ac{MIMO} channels with per-antenna power constraints, we show and compare: i) all-digital solution with power allocation performed using a joint space-frequency waterfilling algorithm with a total power constraint (All-dig TPC), ii) all-digital solution with per-antenna power constraints (All-dig PPC), iii) hybrid solution with per-antenna power constraints (Hybrid PPC), and iv) hybrid solution with per-antenna power constraints using imperfect channel estimates. For the latter, estimates $\{\hat{\bsfH}[k]\}_{k=0}^{K-1}$ are obtained by using the \ac{SS-SW-OMP+Th} algorithm in \cite{RodGonVenHea:TWC_18} (Hybrid PPC SW-OMP), which has been shown to provide near-optimum performance in terms of both estimation error and spectral efficiency \cite{RodGonVenHea:TWC_18}. For the latter, the number of subcarriers processed $K_\text{p}$ and threshold $\beta$ are set to $K_\text{p} = 64$ and $\beta = 0.025\sigma^2$ (see \cite{RodGonVenHea:TWC_18}).

The typical parameters for our system configuration are summarized as follows. Both the transmitter and the receiver are assumed to use Uniform Linear Arrays (ULAs) with half-wavelength separation. Such a ULA has steering vectors obeying the expressions $\{\bsfa_\text{T}(\theta_\ell)\}_n = \sqrt{\frac{1}{\Nt}}e^{\sfj n \pi \cos(\theta_\ell)}$, $n = 0,\ldots,\Nt-1$, and $\{\bsfa_\text{R}(\phi_\ell)\}_m = \sqrt{\frac{1}{\Nr}} e^{\sfj m \pi \cos(\phi_\ell)}$, $m = 0,\ldots,\Nr - 1$. We consider two different setups as:
\begin{itemize}
\item System I: the number of transmit and receive antennas are $\Nt = 64$, $\Nr = 32$, and the number of RF chains is set to $\Lt = 4$, $\Lr = 4$.
\item System II: the number of transmit and receive antennas are $\Nt = 64$, $\Nr = 16$, and the number of RF chains is set to $\Lt = 4$, $\Lr = 2$.
\end{itemize} 
The phase-shifters used in both the transmitter and the receiver are assumed to have $N_\text{Q,Tx}$ and $N_\text{Q,Rx}$ quantization bits for the discrete phases that can be implemented. Therefore, the entries of the analog precoders and combiners belong to the discrete sets ${\cal A}_\text{Tx} = \left\{0,\frac{2\pi}{2^{N_\text{Q,Tx}}},\ldots,\frac{2\pi (2^{N_\text{Q,Tx}}-1)}{2^{N_\text{Q,Tx}}}\right\}$ and ${\cal A}_\text{Rx} = \left\{0,\frac{2\pi}{2^{N_\text{Q,Rx}}},\ldots,\frac{2\pi (2^{N_\text{Q,Rx}}-1)}{2^{N_\text{Q,Rx}}}\right\}$. The number of quantization bits is set to $N_\text{Q,Tx} = N_\text{Q,Rx} = 4$. The number of OFDM sucarriers is set to $K = 256$, and a Zero-Prefix (ZP) of length $Z_\text{P} = 64$ is assumed to remove Inter Symbol Interference (ISI). The \ac{mmWave} frequency-selective channel is generated according to \eqref{eqn:channel_model} using small-scale fading parameters directly generated from QuaDRiGa channel simulator \cite{QuaDRiGa_IEEE}, \cite{QuaDRiGa_Tech_Rep} using the 3GPP TR 38.901 Urban Macrocell (UMa) scenario specified in \cite{5G_channel_model}. We also illustrate the dependency of the numerical results on the Rician factor of the \ac{MIMO} channel.

\begin{figure*}[ht!]
    \centering
\begin{tabular}{cccc}
{\includegraphics[width=0.5\textwidth]{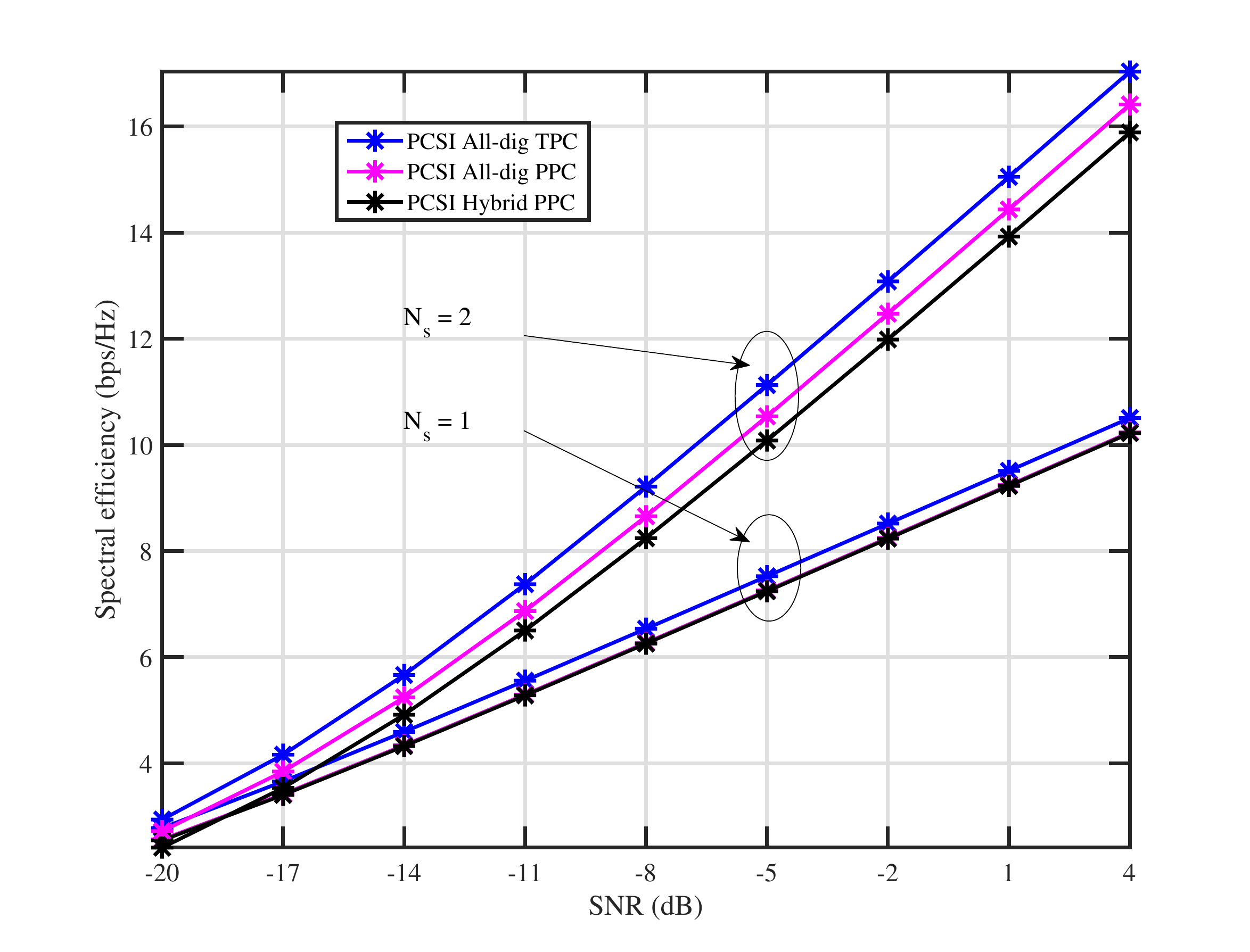}} & {\includegraphics[width=0.5\textwidth]{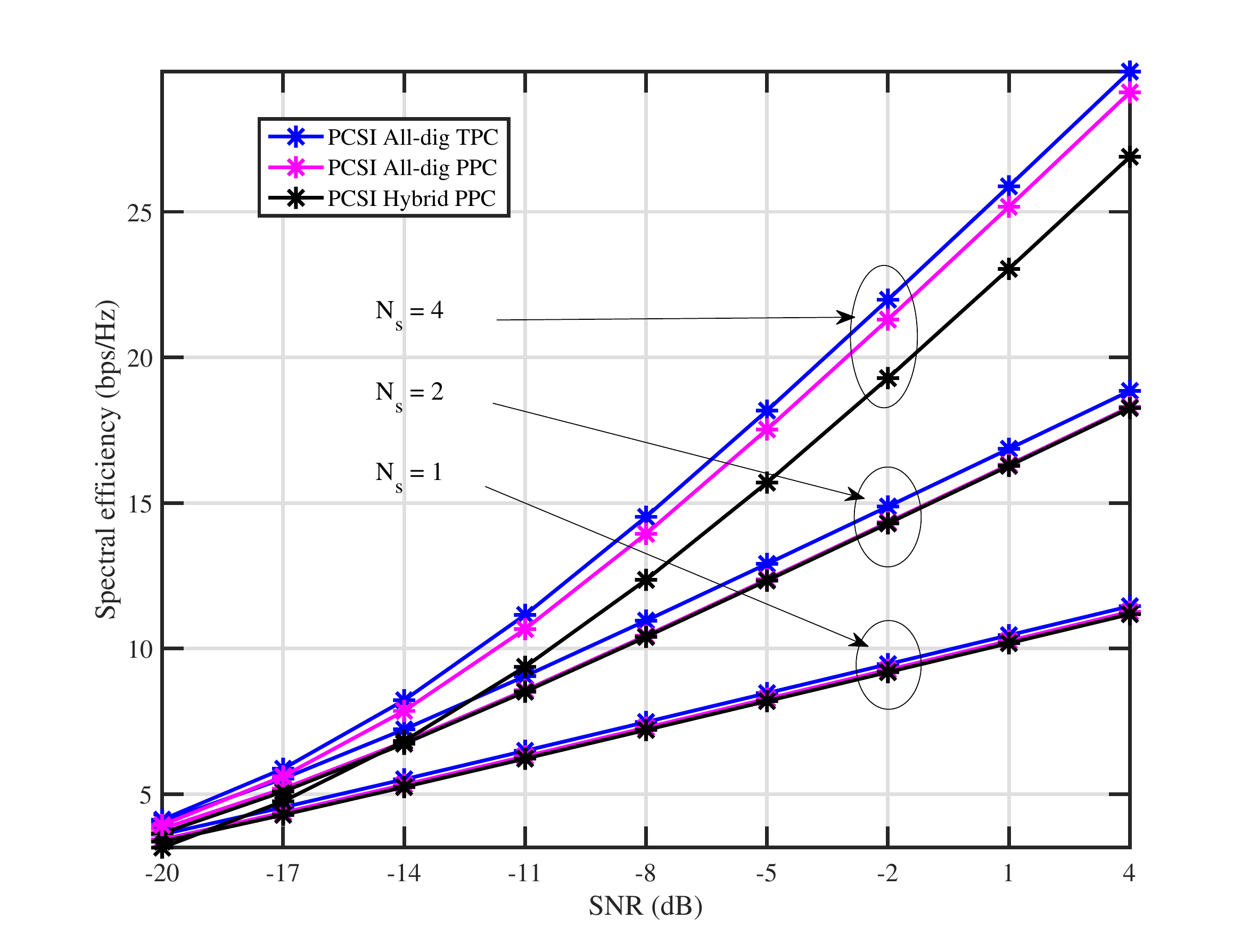}} \\
(a) & (b) \\
{\includegraphics[width=0.5\textwidth]{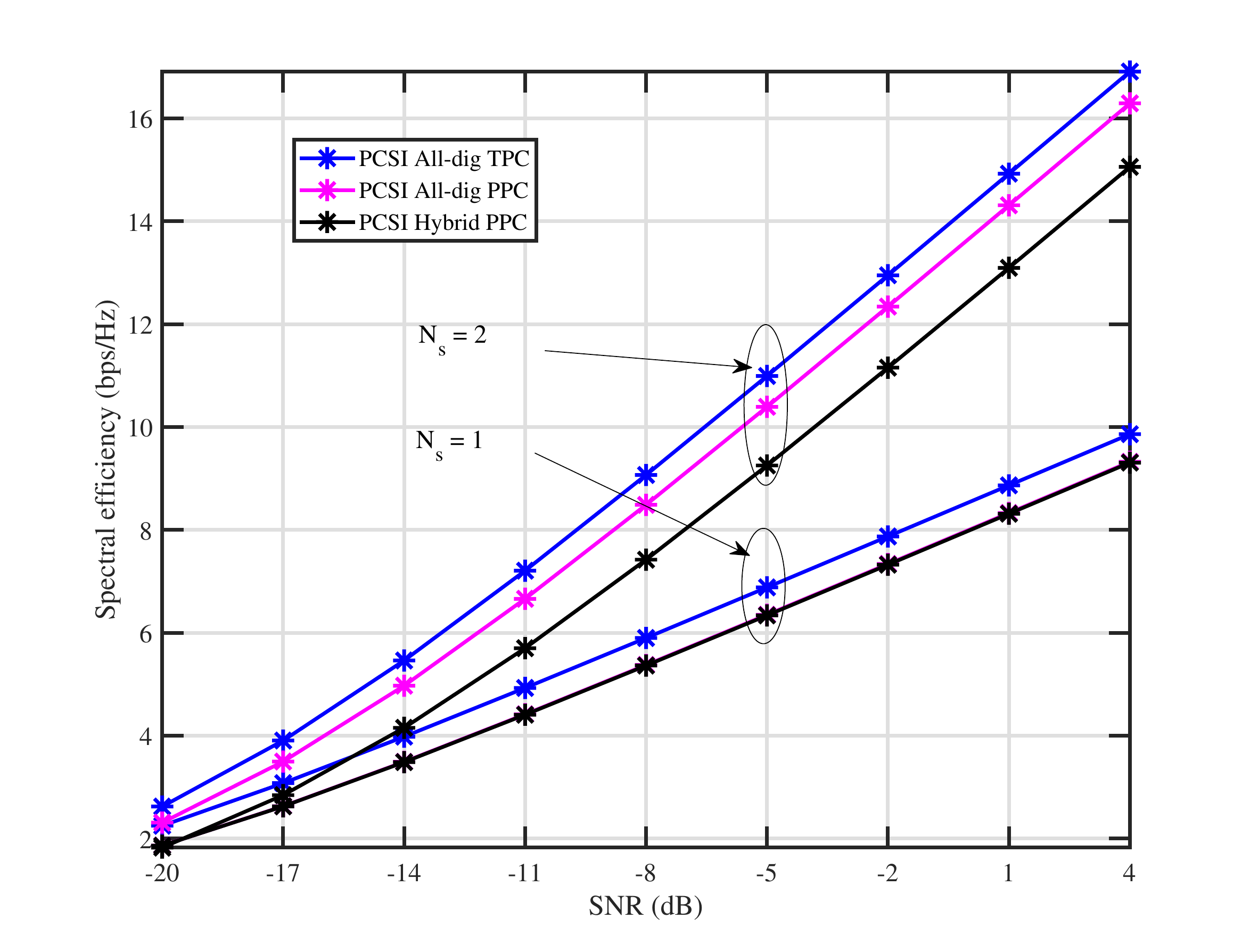}} & {\includegraphics[width=0.5\textwidth]{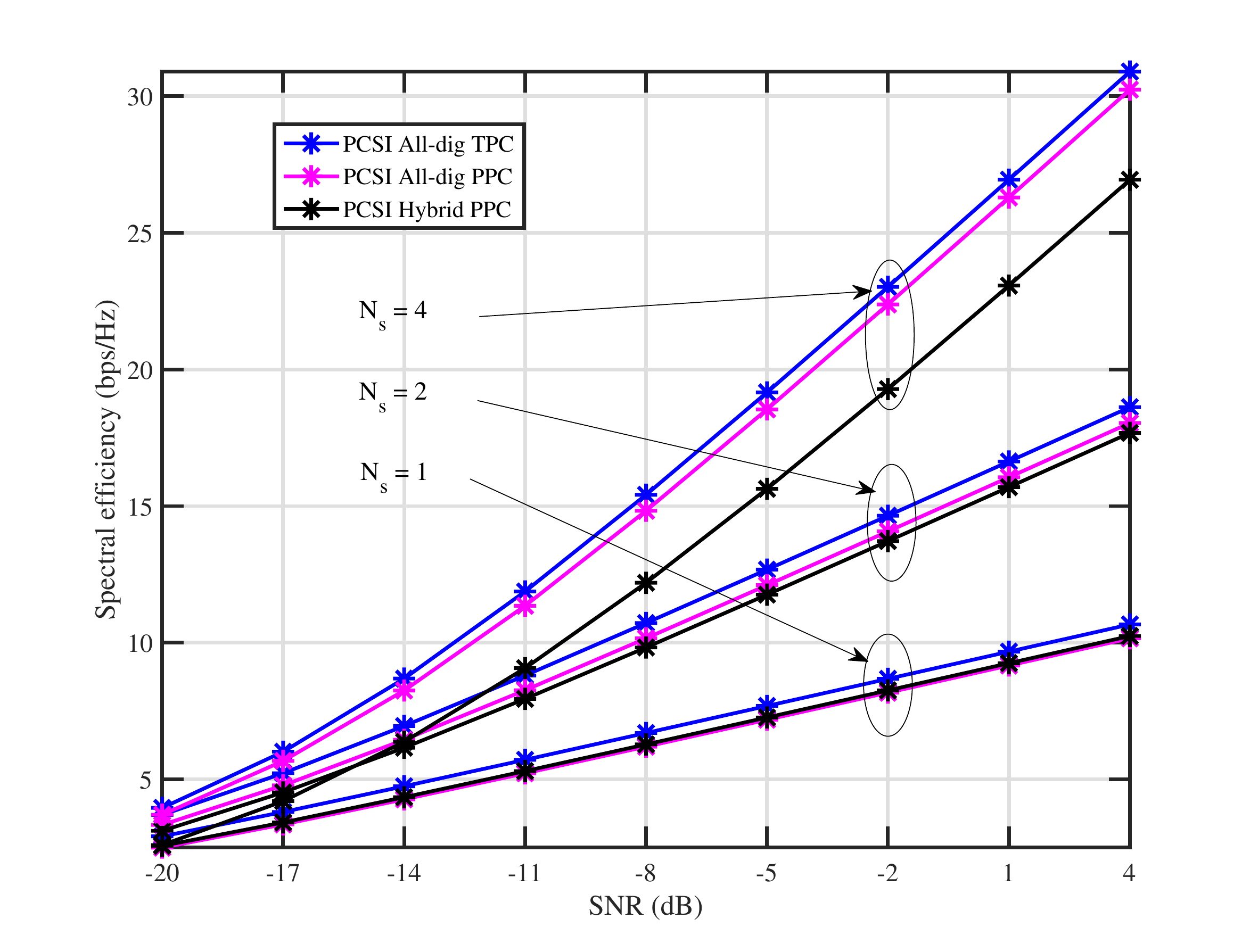}} \\
(c) & (d) \\
    \end{tabular}
    \caption{Comparison of evolution of the spectral efficiency versus $\SNR$ for the different precoding and combining strategies. Figures (a) and (c) are obtained using System I for a Rician factor of $0$ dB (a) and $-10$ dB (c). Figures (b) and (d) are obtained using System II for a Rician factor of $-10$ dB.}.
    \label{fig:SE_vs_SNR_PCSI}
\end{figure*}

We show in Fig. \ref{fig:SE_vs_SNR_PCSI} the average spectral efficiency for the different precoding and combining algorithms, in which System I and System II are compared for two different Rician factors, $\text{KF} = \{-10,0\}$ dB. It can be observed that the performance gap of the All-dig PPC design with respect to the All-dig TPC design method is small, although it increases with the number of transmitted data streams. This effect is driven by two facts: i) the dependency between the singular vectors and singular values of the all-digital precoder $\bsfF[k] = \bsfU_F[k] \bm \Sigma_F[k] \bsfV_F^*[k]$, and most importantly ii) our considering per-antenna power constraints, rather than a total power constraint. 

On the one hand, since we assumed that $\bsfU_F[k]$ and $\bm \Sigma_F[k]$ are decoupled to find closed-form solutions for the original design problem (recall that setting $\bsfV_F[k] = \bI_{\Ns}$ preserves optimality in terms of spectral efficiency), the dependency between these matrices brings about some performance loss as $\Ns$ increases. Notice, however, that this loss is smaller than approximately $0.5$ bps/Hz from Fig. \ref{fig:SE_vs_SNR_PCSI}, and it does not depend on the $\SNR$. We can also observe that this performance loss is small for both System I and System II and the two Rician factors considered. For the narrowband scenario, it was shown in \cite{Globecom_16_per_antenna} that the all-digital PPC design proposed therein performs well even for channel models with several clusters, rays per cluster, and non-negligible angular spread. It is not surprising to conclude, therefore, that the proposed all-digital PPC design is a both reasonable and insightful solution that can be applicable to 5G \ac{mmWave} \ac{MIMO} systems.

On the other hand, considering per-antenna power constraints further limits the achievable spectral efficiency with respect to a system in which a total power constraint is considered. Since a design with per-antenna constraints results in $\Nt+K\Ns$ inequality constraints (including non-negativity of the singular values of $\bm \Sigma_F[k]$), the geometrical set in which we are searching for the power allocation coefficients is further reduced with respect to the event of considering a total power constraint comprising of $K\Ns + 1$ inequality constraints.

 Further, the performance gap between the Hybrid PPC and All-digital PPC designs is small, although it increases with $\Ns$, and it is also affected by the Rician factor. We observe that, for the smallest Rician factor of $-10$ dB, the spectral efficiency loss between the all-digital designs and the Hybrid PPC design is larger, yet small. As $\text{KF} \to \infty$, the channel energy is concentrated on a single multipath component (i.e. the \ac{LoS} component). Consequently, the effective \textit{channel subspace} that the hybrid precoding and combining spatial filters must approximate has minimum dimension, and this leads to more accurate hybrid approximations to the input all-digital precoders and combiners. Conversely, as the Rician factor decreases, the channel energy is more spread amongst the $\sum_{c=1}^{C}R_c$ spatial components, so that the effective \textit{channel subspace} that hybrid precoders and combiners should approximate has maximum dimension, and the $\Lt$ ($\Lr$) available degrees of freedom at transmitter (receiver) are not enough to find a sufficiently accurate hybrid solution exhibiting small enough average chordal distance with the all-digital solutions.
 
 \begin{figure*}[ht!]
    \centering
\begin{tabular}{cccc}
{\includegraphics[width=0.5\textwidth]{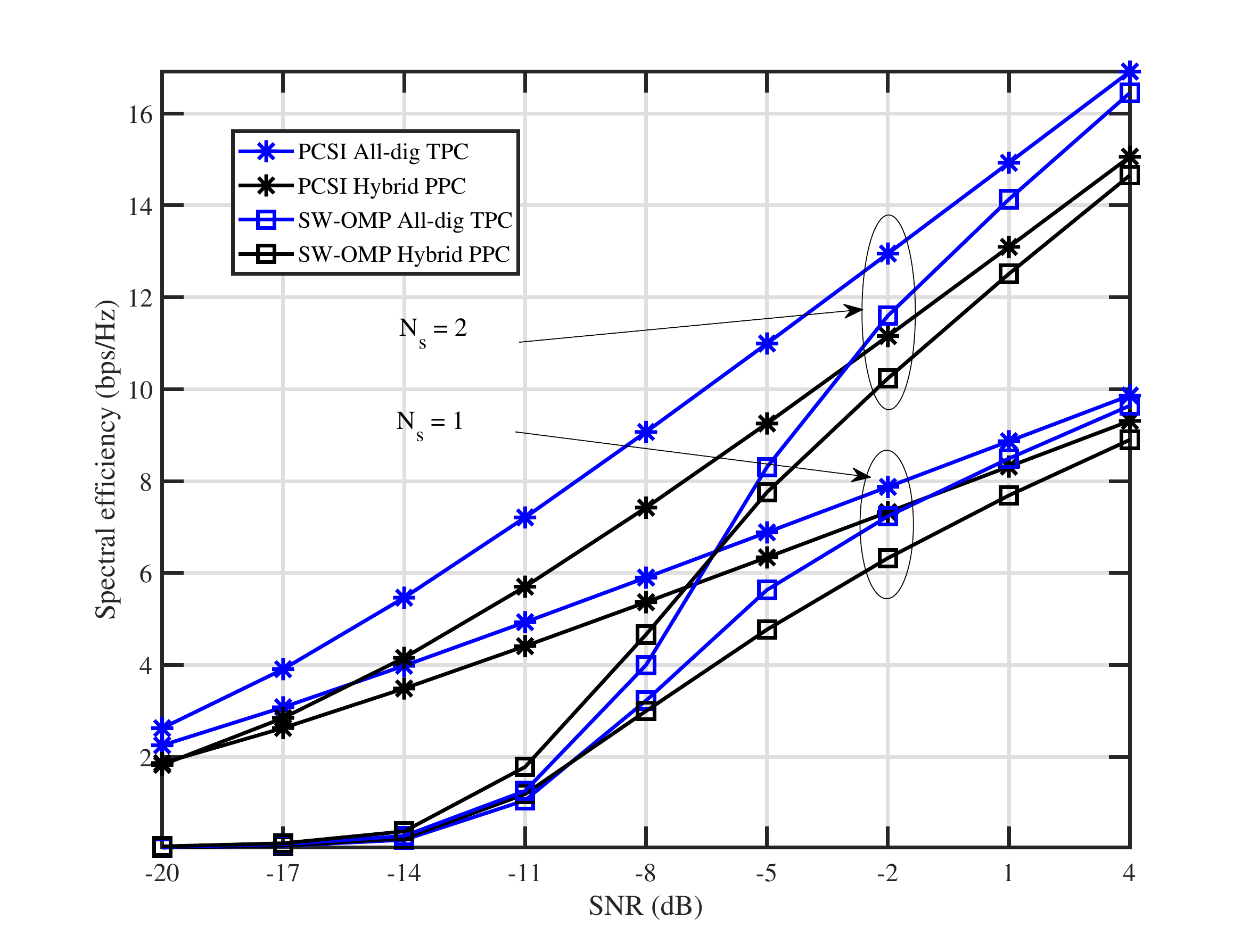}} & {\includegraphics[width=0.5\textwidth]{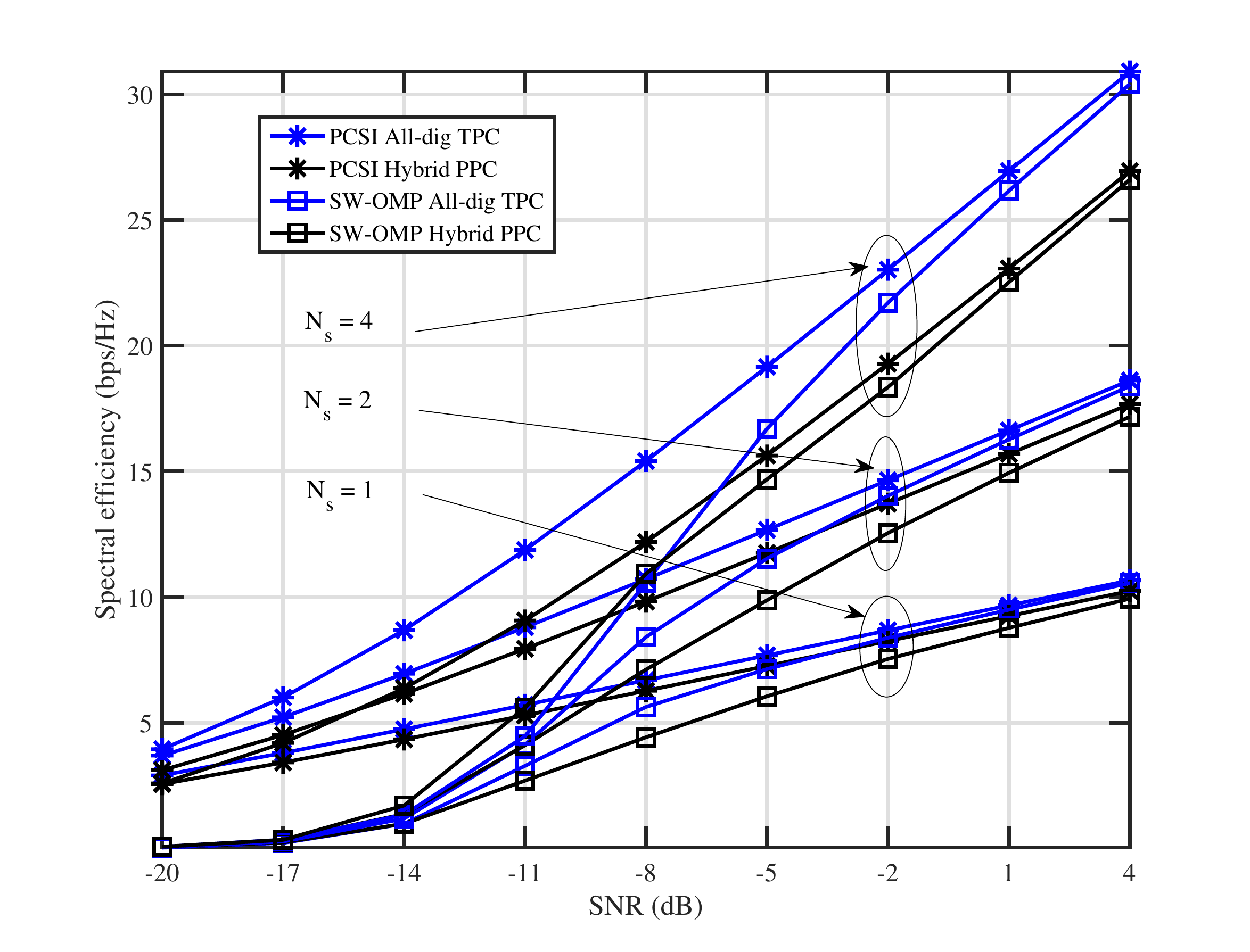}} \\
(a) & (b) \\
    \end{tabular}
    \caption{Comparison of evolution of the spectral efficiency versus $\SNR$ for the different precoding and combining strategies. Figure (a) is obtained using System I, and Figure (b) is obtained using System II, both for a Rician factor of $-10$ dB.}.
    \label{fig:SE_vs_SNR_ICSI}
\end{figure*}

We show in Fig. \ref{fig:SE_vs_SNR_ICSI} the evolution of spectral efficiency versus $\SNR$ for the all-digital TPC and hybrid PPC designs with both perfect CSI and channel estimates obtained using the \ac{SS-SW-OMP+Th} algorithm in \cite{RodGonVenHea:TWC_18} in the absence of beam-squint. We consider a number of $M = 60$ training OFDM symbols to estimate the channel, and angular grids of sizes $\Gt = \Gr = 128$ to promote sparsity in the angular domain. We observe from Fig. \ref{fig:SE_vs_SNR_ICSI}(a) and Fig. \ref{fig:SE_vs_SNR_ICSI} (c) that the performance gap between both all-digital and hybrid channel estimates and their perfect CSI counterparts is not negligible, especially for very low values of $\SNR$. This effect is due to the higher energy spread among the different multipath components as $\text{KF}$ decreases, which leads to a large number of effective multipath components in the estimated channel owing to smaller Fisher Information when $\sum_{c=1}^{C}R_c$ grows, as discussed in \cite{RodGonVenHea:TWC_18}. Therefore, the estimation error of the channel gains also reduces, and results in the algorithm's finding worse estimates of the singular values of $\{\bsfH[k]\}_{k=0}^{K-1}$. Consequently, this leads to lower accuracy when finding the power allocation coefficients for both the TPC and PPC designs. Nonetheless, as discussed in \cite{RodGonVenHea:TWC_18}, this issue can be easily circumvented by slightly increasing the training overhead up to $80-100$ OFDM symbols. However, as analyzed in \cite{RodGonVenHea:TWC_18}, we may conclude that the robustness of the \ac{SS-SW-OMP+Th} algorithm makes it possible to attain near-optimum values of spectral efficiency, especially when $\SNR$ increases.


 \begin{figure*}[ht!]
    \centering
\begin{tabular}{cccc}
{\includegraphics[width=0.5\textwidth]{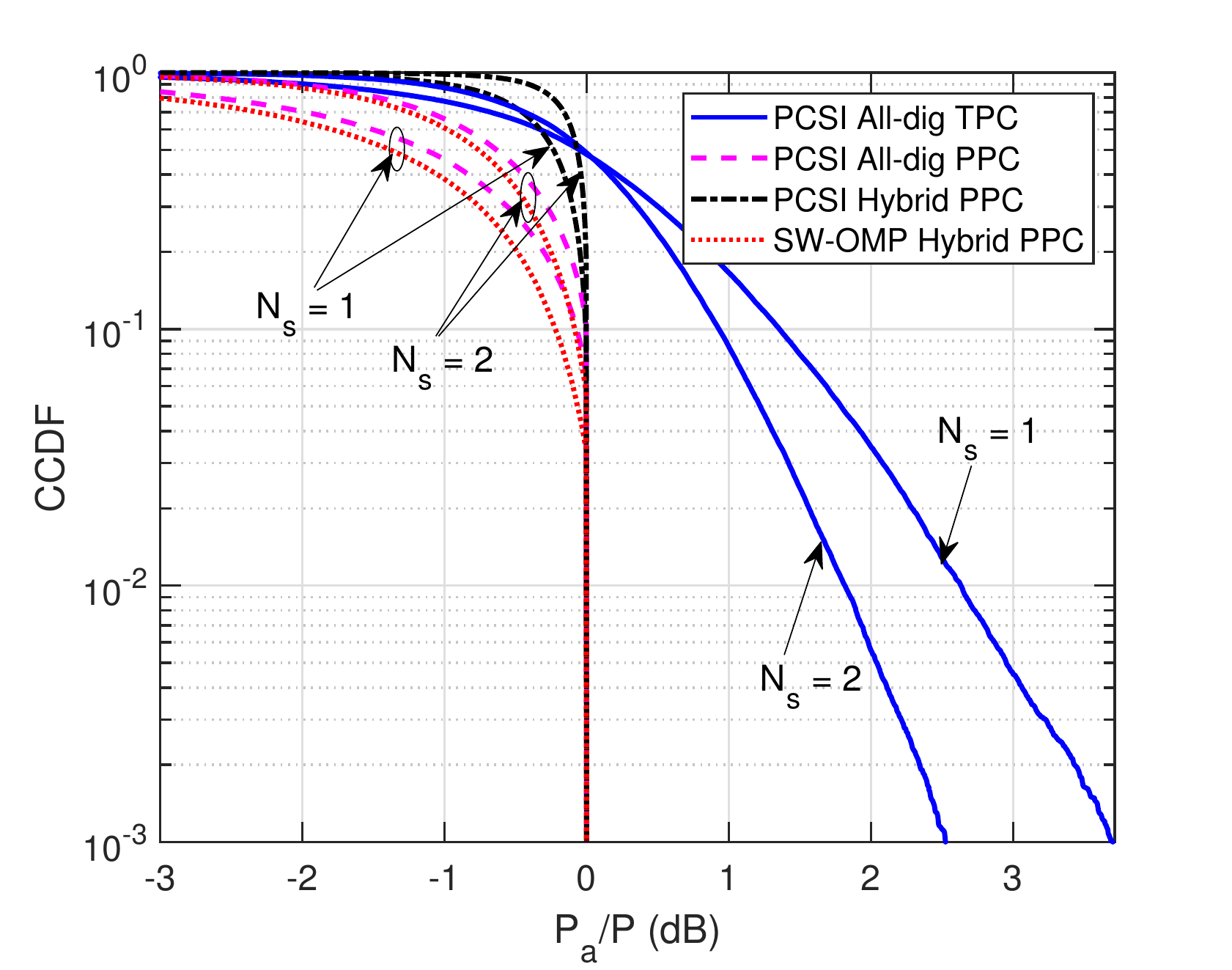}} & {\includegraphics[width=0.5\textwidth]{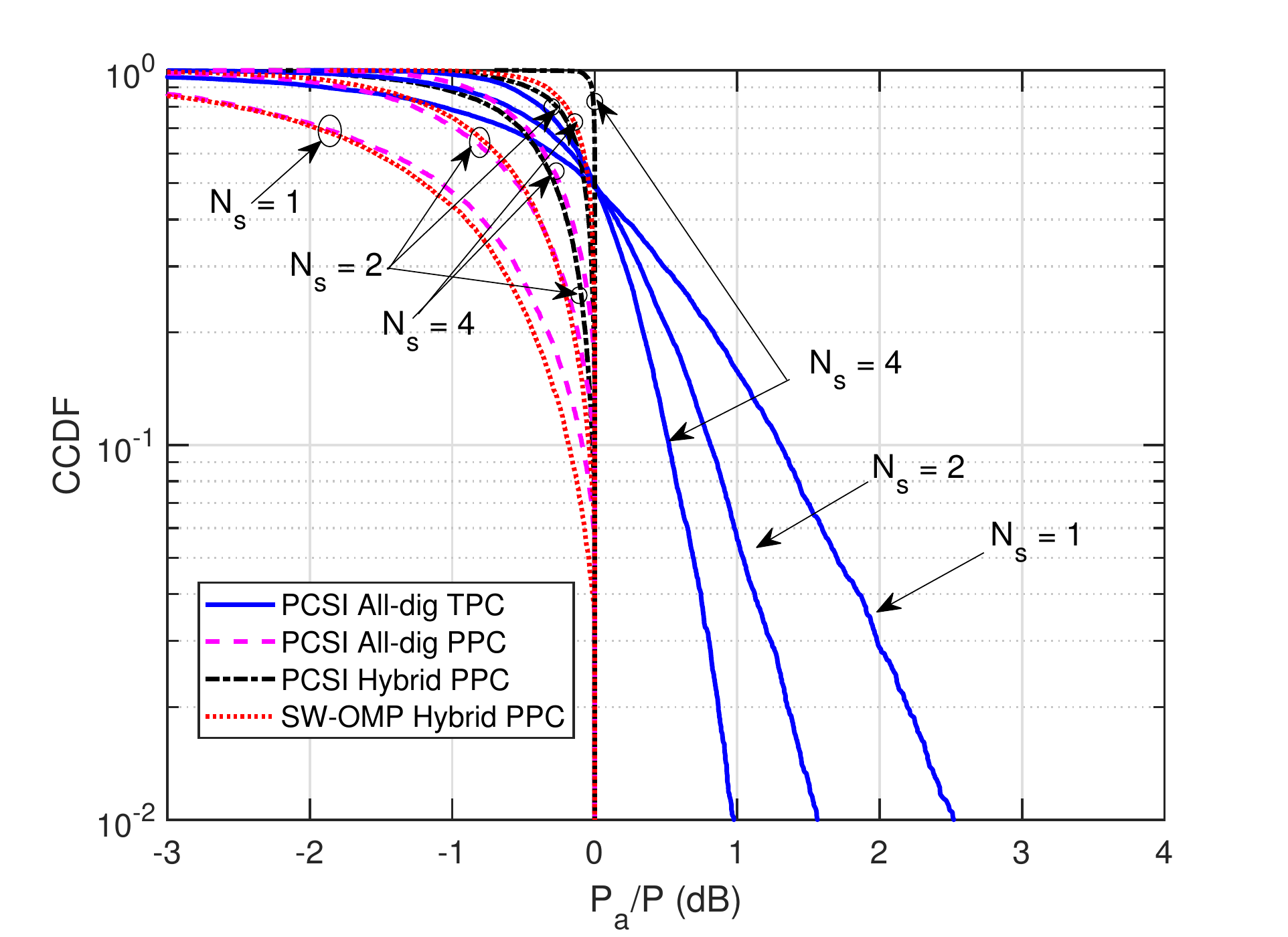}} \\
(a) & (b) \\
    \end{tabular}
    \caption{Comparison of sample \ac{CCDF}s for the different precoding and combining strategies. Figures (a) and (c) are obtained using System I for a Rician factor of $0$ dB (a) and $-10$ dB (c). Figures (b) and (d) are obtained using System II for a Rician factor of $-10$ dB. A number of $N_\text{MC} = 1000$ Monte Carlo channel realizations is considered to obtain these curves.}.
    \label{fig:CCDF}
\end{figure*}

In Fig. \ref{fig:CCDF}, we show the sample \ac{CCDF} of the different precoding and combining strategies for both System I (a) and System II (b), for a \ac{mmWave} \ac{MIMO} channel with Rician factor $-10$ dB. We observe that, for both the all-digital and hybrid PPC designs, the per-antenna constraints are always met, yet not necessarily with equality. We notice, however, that as $\Ns$ increases, the probability that the power delivered to the $j$-th antenna is exactly equal to $p_j$ increases. In fact, for the hybrid PPC design with perfect CSI and $\Ns = 4$ in Fig. \ref{fig:CCDF} (b), the per-antenna constraints are seen to be met with equality with probability of almost $1$ . We observe that the hybrid PPC design yields the largest delivered power to any given antenna, which are partially due to the hardware constraints that the hybrid \ac{MIMO} architecture imposes to the hybrid design. When projecting the designed all-digital precoder onto the feasible set of matrix with unit-modulus entries, this reduces orthogonality with respect to the all-digital solution. Due to this, a higher amount of power is allocated in the different data streams to compensate for the energy loss coming from this orthogonality mismatch, thereby yielding larger values for the power delivered to any given antenna. Further, we observe that the all-digital TPC design does not meet per-antenna power constraints, and results in larger power spread as $\Ns$ decreases. From Fig. \ref{fig:CCDF} (a), we observe that, if a probability of meeting the per-antenna constraints of $10^{-3}$ is required, the input power would have to be backed off by approximately $2.5$ and $3.5$ dB for the all-digital TPC design. Thereby, the spectral efficiency curves in Fig. \ref{fig:SE_vs_SNR_PCSI}(a) and Fig. \ref{fig:SE_vs_SNR_PCSI} (c) would need to be shifted to the right by the same amount, leading to our proposed PPC design exhibiting higher effective spectral efficiency. This clearly shows the importance of considering per-antenna power constraints when a practical \ac{mmWave} \ac{MIMO} system is designed.

 \begin{figure*}[ht!]
    \centering
\begin{tabular}{cccc}
{\includegraphics[width=0.5\textwidth]{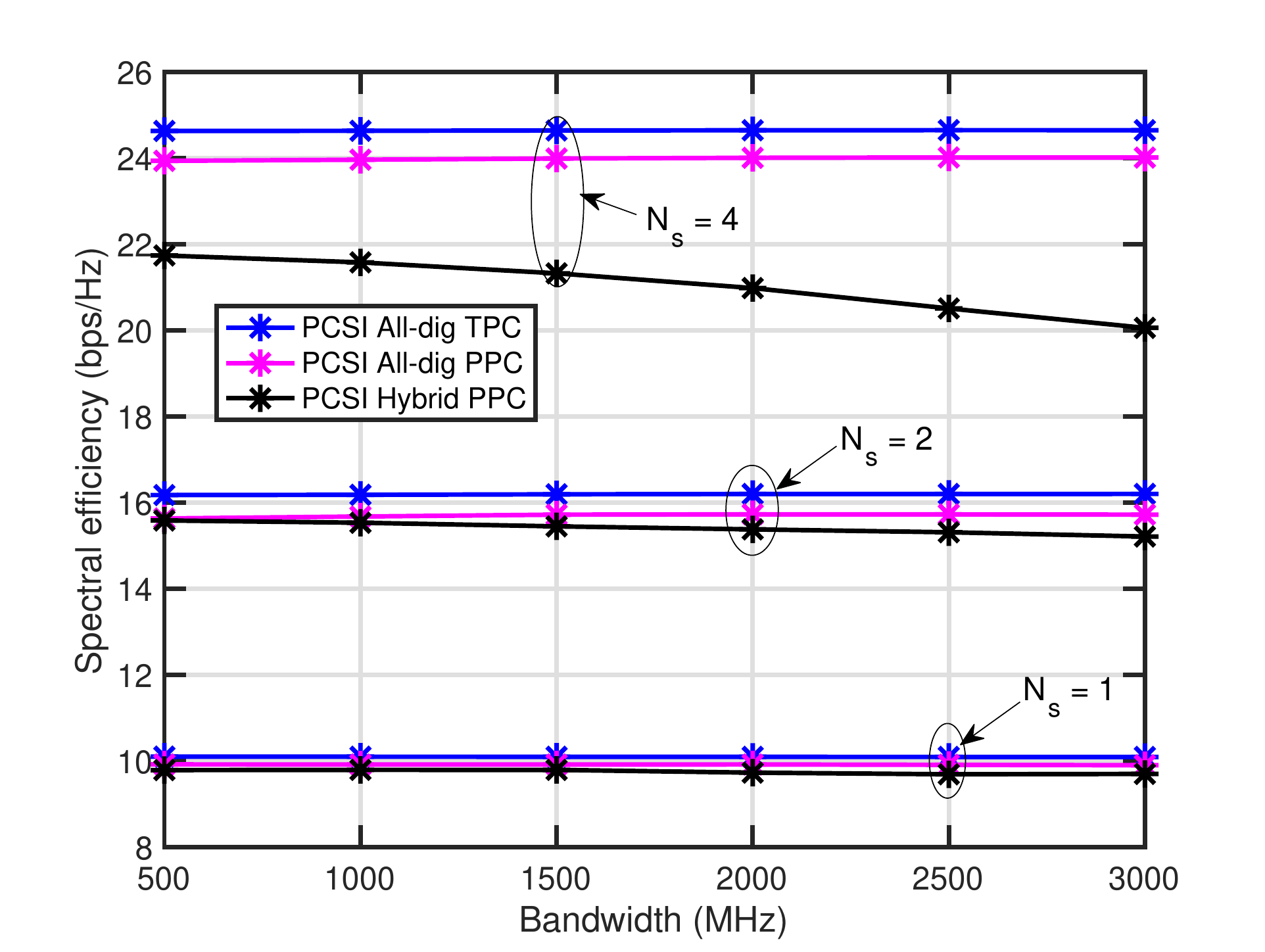}} & {\includegraphics[width=0.5\textwidth]{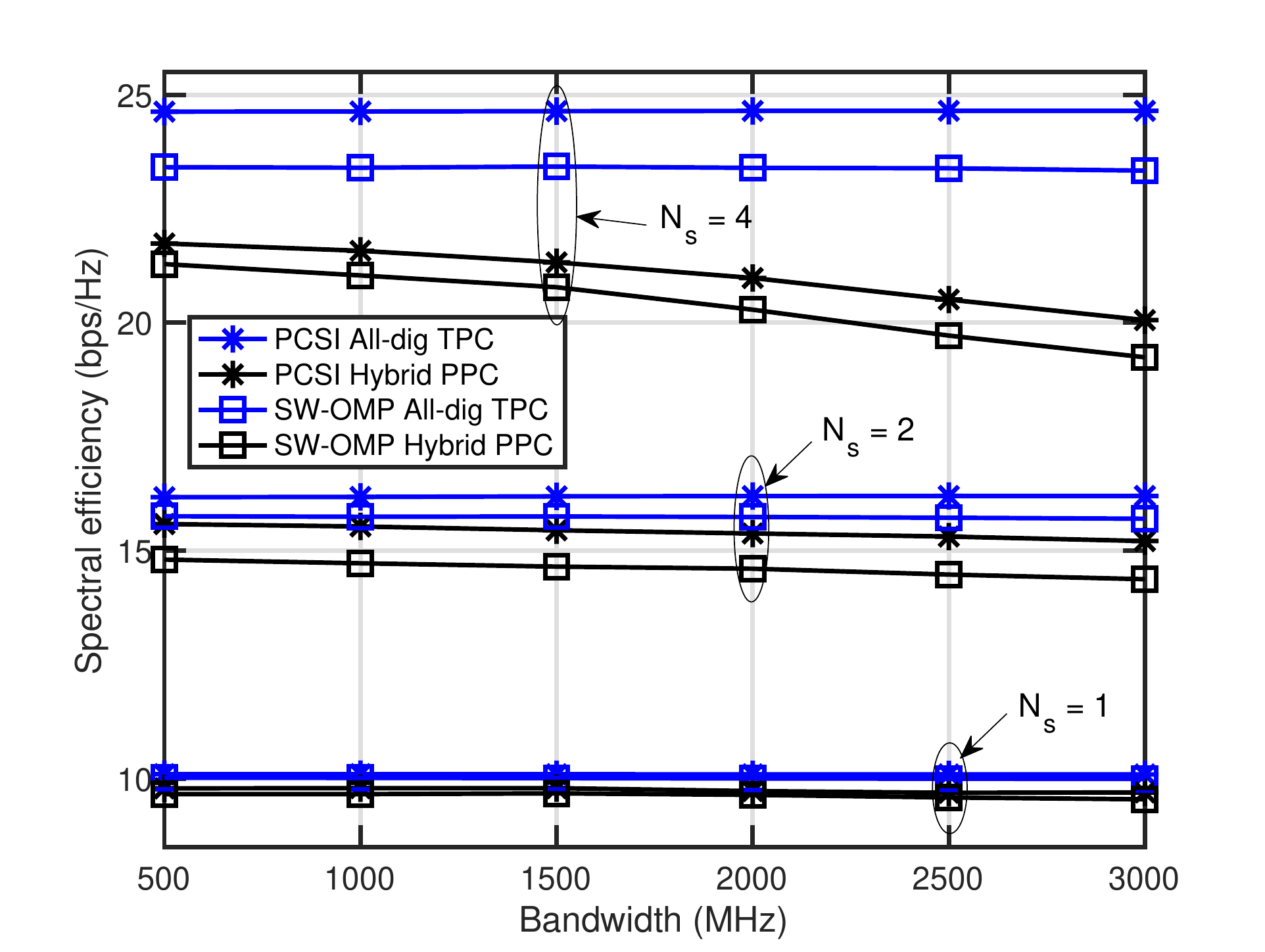}} \\
(a) & (b) \\
    \end{tabular}
    \caption{Comparison of evolution of the spectral efficiency versus $B$ for the different precoding and combining strategies, for both perfect CSI (a) and channel estimates obtained using the \ac{SS-SW-OMP+Th} algorithm from \cite{Globecom_18_beam_squint}. The Rician factor is set to $0$ dB.}.
    \label{fig:SE_vs_B}
\end{figure*}

 \begin{figure*}[ht!]
    \centering
\begin{tabular}{cccc}
{\includegraphics[width=0.5\textwidth]{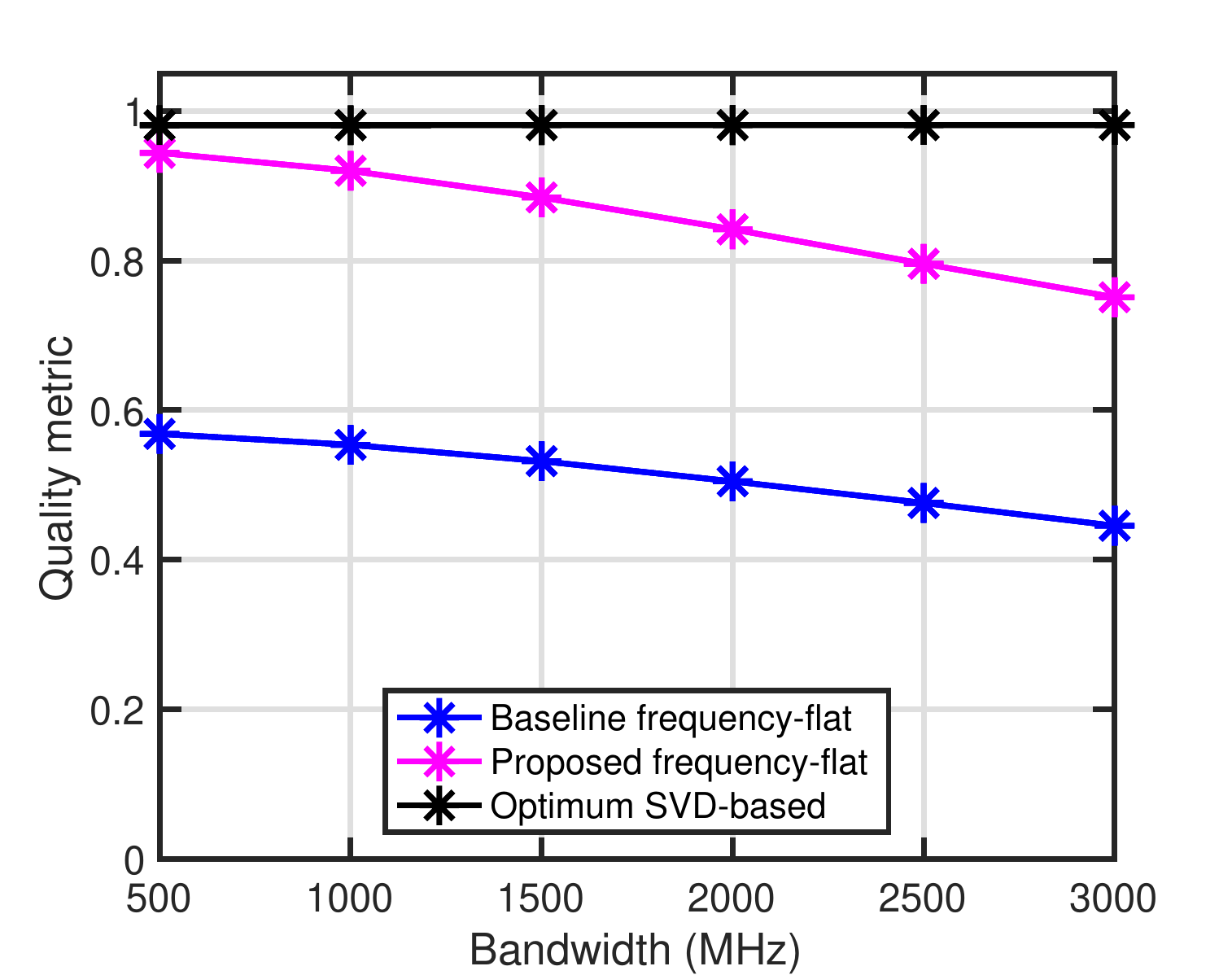}} & {\includegraphics[width=0.5\textwidth]{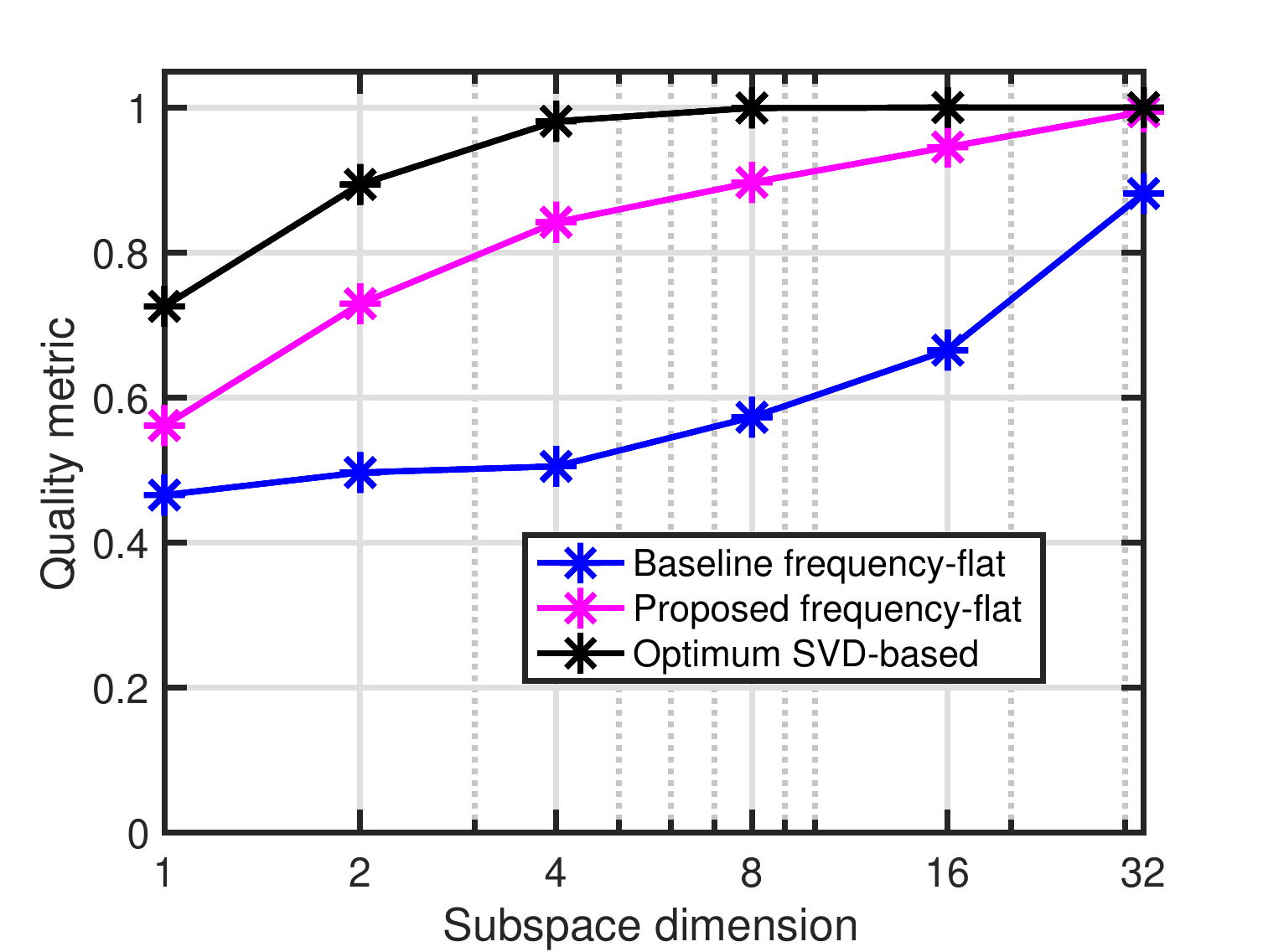}} \\
(a) & (b) \\
    \end{tabular}
    \caption{Comparison of evolution of the quality metric in \eqref{equation:Quality_subspaces} versus $B$ (a) and the subspace dimension $d$ (b). The Rician factor is set to $0$ dB.}.
    \label{fig:Quality_subspaces}
\end{figure*}

Last, we also analyze the impact of beam-squint on the design of the frequency-selective hybrid precoders and combiners, for both perfect CSI and channel estimates. For the latter, we consider the modified \ac{SS-SW-OMP+Th} algorithm in \cite{Globecom_18_beam_squint}, for which the same parameters as with Fig. \ref{fig:SE_vs_SNR_ICSI} are selected to run the estimation algorithm. Furthermore, the number of groups of subcarriers is set to $G = 4$, whilst the number of subcarriers per group is set to $K_\text{g} = K/G = 16$, and a total of $K_\text{p} = K_\text{g}$ subcarriers are exploited to compute the support of the quasi-sparse \ac{mmWave} \ac{MIMO} channel. We show in Fig. \ref{fig:SE_vs_B} (a) and Fig. \ref{fig:SE_vs_B} (b) the average ergodic spectral efficiency as a function of the bandwidth $B$ for $\Ns = \{1,2,4\}$, for an average $\SNR = 0$ dB and Rice factor of $0$ dB under System I, considering both perfect (a) and imperfect CSI (b). We observe that, as the bandwidth increases, the channel's left and right singular basis are less related to each other, for the different subcarriers. Consequently, the all-digital matrix $\bsfT$ in Section \ref{sec:hybrid_precoders} (see \eqref{equation:subopt_RF_precoder}) has higher rank, and it becomes harder to fully represent the different frequency-domain subchannels with a single hardware-constrained RF precoding matrix. Thereby, spectral efficiency decreases as the bandwidth increases, as shown in Fig. \ref{fig:SE_vs_B} (a) and (b). Further, we also observe that the channel estimation algorithm in \cite{Globecom_18_beam_squint} is able to find reliable enough channel estimates and therefore achieve near-optimum values of spectral efficiency, as shown in Fig. \ref{fig:SE_vs_B} (b). 

To illustrate the relationship between the channel's singular subspaces, we consider the following quality metric:
\begin{equation}
	\gamma = \frac{\sum_{k=0}^{K-1}\left\|\left[\bsfU_S\right]_{:,1:d}^* \bsfH[k] \left[\bsfU_T\right]_{:,1:d}\right\|_F^2}{\sum_{k=0}^{K-1}\left\|\bsfH[k]\right\|_F^2}
	\label{equation:Quality_subspaces}
\end{equation}
The metric in \eqref{equation:Quality_subspaces} comprises of both a straightforward and insightful approach to assess the relationship between the different channel subspaces at different subcarriers. In Fig. \ref{fig:Quality_subspaces} , we show this metric for System I and a Rician factor of $0$ dB, as a function of $B$ (a) and subspace dimension $d$ in \eqref{equation:Quality_subspaces} (b). We also compare the proposed precoding and combining strategy with the approach in \cite{KiranNuriaRobert}, which is based on performing a QR factorization of the transmit and receive arrays matrices, for the subcarrier placed at half the bandwidth of the transmitted signal. As we can observe in Fig. \ref{fig:Quality_subspaces} (a), as the bandwidth increases, the relationship between the channel's singular basis is destroyed and the frequency-flat precoder and combiner need a larger number of degrees of freedom to accurately represent the channel's eigenspaces. This effect is shown in Fig. \ref{fig:Quality_subspaces} (b), wherein it is shown that, as $d$ increases, the frequency-flat precoder and combiner can represent the channel's singular spaces more reliably. Therefore, it is clear that optimizing the chordal distance between the unconstrained all-digital precoders (combiners) and their hybrid counterparts is a reasonable strategy that allows obtaining near-optimum data rates.

\section{Conclusions}

In this paper, we considered the problem of frequency-selective hybrid precoding and combining with per-antenna power constraints, and proposed a novel algorithm to design these spatial filters using hybrid architectures. We theoretically analyzed the problem and provided an upper bound for the spectral efficiency that performs close to the perfect CSI design with a total power constraint. We analyzed the influence of different system parameters on the final achievable performance, both in terms of spectral efficiency and power delivered to any given antenna. Further, we also showed that our proposed design performs well even when the beam-squint effect is considered and the signal bandwidth is as large as $3$ GHz, and provided the subspace-based intuition to understand the effect of beam-squint on the different frequency-domain subchannels. Future work would conduct a theoretical analysis on the effect of beam-squint on hybrid precoders and combiners for other types of hybrid architecture, and extend the proposed framework to the multi-user scenario.


\bibliographystyle{IEEEtran}

\end{document}